\documentclass[amsmath,amssymb,floatfix,prb,preprint,superscriptaddress]{revtex4-1}

\usepackage{graphicx}
\usepackage{physics}
\usepackage{color}
\usepackage{amsmath}
\usepackage{fullpage}
\usepackage{hyperref}
\usepackage{float}
\usepackage{times}
\usepackage{soul}

\usepackage{subcaption}

\hypersetup{
    colorlinks = true,
    allcolors = [rgb]{0,0,0},
}

\begin{document}

\title{Experimental demonstration of quantum effects in the operation of microscopic heat engines}

\author{J. Klatzow}
\affiliation{Clarendon Laboratory, University of Oxford, Parks Road, Oxford OX1 3PU, United Kingdom}
\author{J. N. Becker}
\affiliation{Clarendon Laboratory, University of Oxford, Parks Road, Oxford OX1 3PU, United Kingdom}
\author{P. M. Ledingham}
\affiliation{Clarendon Laboratory, University of Oxford, Parks Road, Oxford OX1 3PU, United Kingdom}
\author{C. Weinzetl}
\affiliation{Clarendon Laboratory, University of Oxford, Parks Road, Oxford OX1 3PU, United Kingdom}
\author{K.T. Kaczmarek}
\affiliation{Clarendon Laboratory, University of Oxford, Parks Road, Oxford OX1 3PU, United Kingdom}
\author{D. J. Saunders}
\affiliation{Clarendon Laboratory, University of Oxford, Parks Road, Oxford OX1 3PU, United Kingdom}
\author{J. Nunn}
\affiliation{Centre for Photonics and Photonic Materials, Department of Physics, University of Bath, Claverton Down, Bath BA2 7AY, United Kingdom}
\author{I. A. Walmsley}
\affiliation{Clarendon Laboratory, University of Oxford, Parks Road, Oxford OX1 3PU, United Kingdom}
\author{R. Uzdin}
\affiliation {Schulich Faculty of Chemistry, Technion -- Israel Institute of Technology, Haifa 3200000, Israel}
\author{E. Poem}
\affiliation{Clarendon Laboratory, University of Oxford, Parks Road, Oxford OX1 3PU, United Kingdom}
\affiliation {Department of Physics of Complex Systems, Weizmann Institute of Science, Rehovot 7610001, Israel}

\maketitle

\begin{figure}[H]
\begin{centering}
\includegraphics[width = 0.65\textwidth]{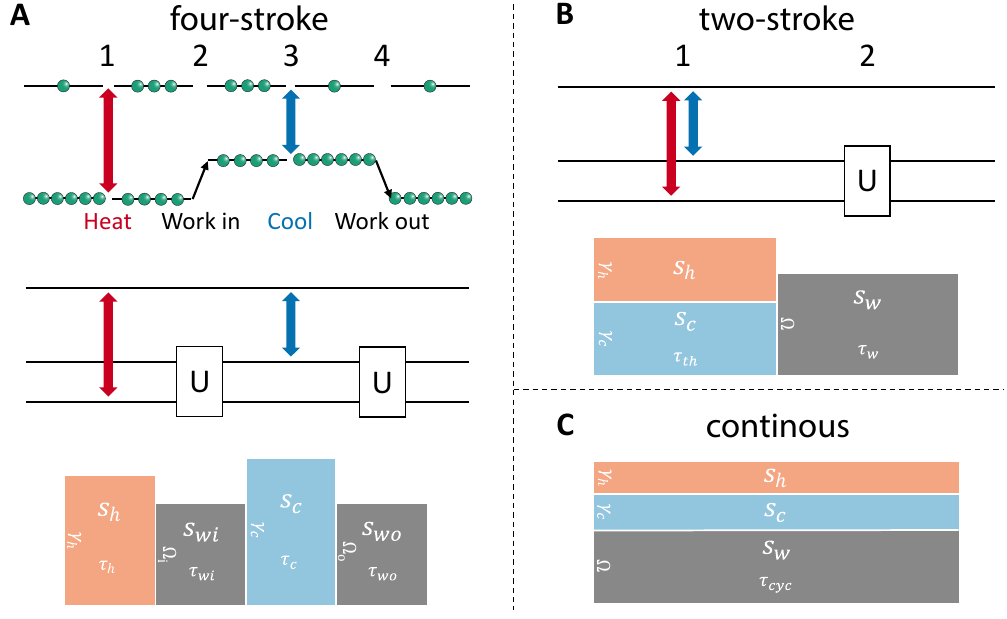}
\caption{\label{fig1} \textbf{Quantum heat engine schematics. }{\bf a,} An example of a four-stroke engine. Top: The classical case of two levels with a variable energy gap. Population is transferred by thermal couplings to a cold bath (blue) and a hot bath (red). Coherence has no role in this case. Middle: A generalization by embedding the operation into three constant levels with general unitary transformations performed on the lowest two levels, and the cold (hot) bath couplings on the top two (upper most and lower most) levels. A swap operation (a $\pi$-pulse) would correspond to the classical case of the top panel. Bottom: schematic description. The gray rectangles represent the unitary (work-extracting) operations, while the orange (light blue) rectangle represents coupling to a hot (cold) thermal bath. The horizontal dimension represents time, while the vertical represents coupling rate. The area of each rectangle is the stroke action (in units of $\hbar$). {\bf b,} The two-stroke engine. Top: three-level embedding. Bottom: Schematic description. {\bf c,} Schematic description of the continuous engine. All the three examples presented here have the same coherent and thermal actions per cycle. In this work we experimentally implement the continuous and two-stroke engines.}
\end{centering}
\end{figure}

\textbf{The heat engine, a machine that extracts useful work from thermal sources, is one of the basic theoretical constructs and fundamental applications of classical thermodynamics. The classical description of a heat engine does not include coherence in its microscopic degrees of freedom. By contrast, a \emph{quantum} heat engine might possess coherence between its internal states. Although the Carnot efficiency cannot be surpassed\cite{Goold2015review,SaiJanetReview,millen2015review}, and coherence can be performance degrading in certain conditions\cite{k176,k215,plastina2014irreversible,Seifert2016PeriodicEngine,Seifert2017,Deng2017}, it was recently predicted that even when using only thermal resources, internal coherence can enable a quantum heat engine to produce more power than \emph{any} classical heat engine using the same resources\cite{EquivPRX,RUcollective}. Such a power boost therefore constitutes a quantum thermodynamic signature.
It has also been shown that the presence of coherence results in the thermodynamic equivalence of different quantum heat engine types\cite{EquivPRX,RUnonMarkovianEquiv}, an effect with no classical counterpart.  
Microscopic heat machines have been recently implemented with trapped ions\cite{rossnagelIonEngExp,Sing3ionEng2017}, and proposals for heat machines using superconducting circuits\cite{PekolaSCengine,campisi2014FT_SolidStateExp} and optomechanics\cite{Kurizki2015workOptoMech,ZhangOptoMechEng}
have been made. When operated with standard thermal baths, however, the machines implemented so far have not demonstrated any inherently quantum feature in their thermodynamic quantities. Here we implement two types of quantum heat engines by use of an ensemble of nitrogen-vacancy centres in diamond, and experimentally demonstrate both the coherence power boost and the equivalence of different heat-engine types. This constitutes the first observation of quantum thermodynamic signatures in heat machines.} 
\clearpage
A quantum heat engine consists of a microscopic system, or an ensemble of such systems, whose internal state can be a coherent superposition of energy states. The engine cycle consists of a sequence of operations (strokes), which include the interaction of the system (or part thereof) either with a thermal bath (cold or hot), or with an external classical/semi-classical field responsible for work extraction. Interactions with the thermal baths act to change the populations of the energy states of the heat engine incoherently, in contrast to the field, which changes the populations coherently. Fig.~\ref{fig1} schematically presents three basic quantum heat-engine types: continuous, two-stroke and four-stroke, which differ by the ordering of the different strokes. Of these types, the four-stroke engine bears the strongest resemblance to macroscopic classical engines such as the Otto engine. It can be described (classically) by a two level system undergoing a four part cycle, illustrated in the top panel of Fig.~\ref{fig1}a, consisting of alternating couplings to the hot and cold baths, interspersed with couplings to the work reservoir, whose effect is to change the spacing between the levels. It can be shown that this dynamics is equivalent to classical swap operations in a multilevel system\cite{EquivPRX} (multilevel embedding), as shown in the middle panel of Fig.~\ref{fig1}a (taking $U = swap$). In a quantum heat engine, the operator U can be any unitary transformation, allowing for the generation of coherence between the two lower levels during the application of the external field. It is coherence generated in this manner that enables quantum heat engines to exhibit non-classical behaviour and so underlies the results in this paper. Note that no non-thermal energy sources such as squeezed baths \cite{LutzSqueezedBaths,ParrondoSqueezedBath,WolgangSqueezedErgotropy},
or externally injected coherence \cite{MitchisonHuber2015CoherenceAssitedCooling,scully2011quantum,Kurizki_MultiatomCoherence} are required. A general four-stroke cycle is schematically illustrated in the bottom panel of Fig.~\ref{fig1}a. One can simplify the cycle structure of the engine by combining the distinct hot and cold strokes into a single thermal stroke, obtaining a two-stroke engine 
(Fig.~\ref{fig1}b). Reducing the complexity of the engine operation even further, one obtains the continuous engine, implemented by simply coupling the system to both thermal baths and the external field continuously (Fig.~\ref{fig1}c). 

The coherence-related improvement in work and power output of a quantum heat engine, constituting a quantum thermal signature (QTS), can be understood as follows: when the coherence produced by the field during the work stroke is not completely erased during the thermal stroke, the work output, proportional to the population change during the work stroke, would be proportional to $C\tau_w$, where $C$ is the surviving coherence and $\tau_w$ is the duration of the work stroke. In contrast, for a fully stochastic engine, where no coherence survives the thermal stroke, the population change is only due to the coherence produced within the same work stroke, and the work output is therefore \textit{quadratic} in $\tau_w$. For a fixed duty cycle $d=\tau_w/\tau_{cyc}$, where $\tau_{cyc}$ is the cycle time, the average power of a stochastic engine is therefore linear in $\tau_{cyc}$, while that of a quantum engine is constant. Thus, for $\tau_{cyc}\rightarrow0$, the power of a stochastic engine vanishes, while that of a quantum engine does not. Note that this picture\cite{EquivPRX} holds only when the \textit{engine action} is significantly less than $\hbar$. The action of a periodic quantum heat engine is formally defined as the operator norm of the generator of motion integrated over the cycle\cite{EquivPRX}. 
For a two-stroke quantum heat engine (See Fig.~\ref{fig1}b) the action is simply given by $s=\hbar\left[\Omega d+\gamma_{th}\left(1-d\right)\right]\tau_{cyc}$, where $\Omega$ is the Rabi frequency, proportional to the amplitude of the external field, and $\gamma_{th}$ is the total coupling rate to the heat baths (including both population transfer and pure dephasing). Thus, for the action to be small enough, both $\Omega$ and $\gamma_{th}$ should be much lower than the repetition rate of the engine. 

In this small-action regime we also observe the predicted quantum heat machine equivalence (QHME)\cite{EquivPRX}: the three basic quantum heat engine types (four-stroke, two-stroke and continuous) all show the same performance per cycle. This effect takes place since in time-symmetric cycles, the strokes commute up to second order in the action. Therefore, for small actions, the order of the strokes does not matter.
%

The system we use in order to experimentally demonstrate QTS and QHME is an ensemble of negatively charged nitrogen vacancy (NV$^-$) centres in diamond\cite{NV_Review}. The NV$^-$ centre is an atomic-like system that exhibits several features desirable for this purpose. 
First, its ground state contains three spin states, $|-1\rangle$, $|0\rangle$, and $|+1\rangle$, that maintain coherence even at room temperature, and can coherently interact with a microwave (MW) field, that can serve as the work reservoir. Second, after optical excitation, the system decays back into the ground-state manifold both by direct, spin-preserving, optical de-excitation (fluorescence), and by spin-dependent non-radiative channels, through a meta-stable spin-singlet state $|0'\rangle$. The system therefore tends to a steady state with a population difference between the different spin components of the ground state. The dynamics of this process is equivalent to that produced by heat bath coupling [see the Supplementary Information (SI)]. 
Finally, the fluorescent decay channel provides a direct means to measure the populations within the ground-state manifold. As spin is preserved during the optical excitation, but the non-radiative decay is spin-dependent, the fluorescence intensity is also spin dependent. The power output of the engine, related to the change in population within the ground-state manifold, can thus be deduced from the change in fluorescence intensity upon introduction of the MW field. This technique, known as optically detected magnetic resonance\cite{NV_Review}, provides a significant advantage over direct measurement of microwave amplification, especially when working in the small action regime. 
%
\begin{figure}[H]
\begin{centering}
\includegraphics[width = 0.65\textwidth]{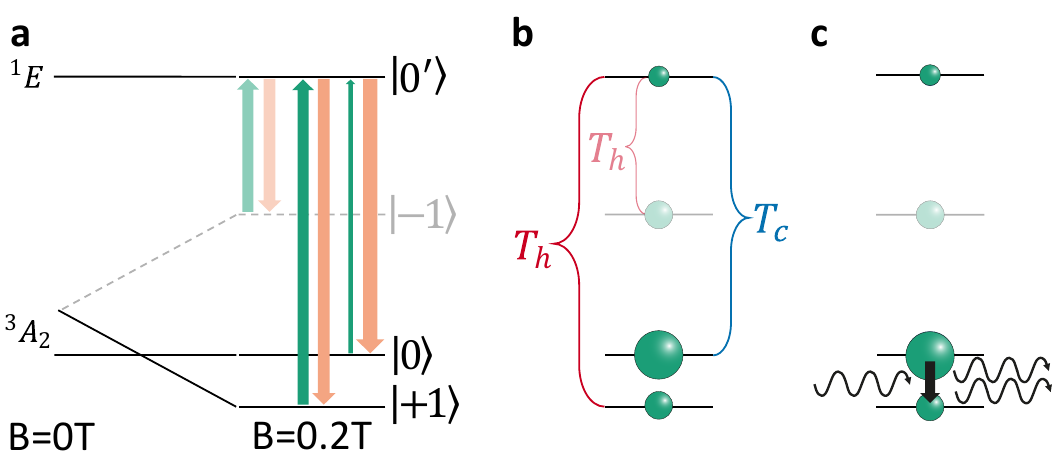}
\caption{\label{fig2} \textbf{Coherent heat engine with NV$^-$ centers in diamond.} {\bf a,} The ground state triplet and the lower intermediate singlet states of the NV$^-$ under axial magnetic field. Optical excitation (green upward arrows) and inter-system crossing (orange downward arrows) transfer population from the $|\pm 1\rangle$ states to the $|0\rangle$ state through the upper $|0'\rangle$ state. The widths of the arrows represent the transition rates. {\bf b,} Steady state populations emulate the equilibrium populations due to coupling of $|0\rangle$ and $|0'\rangle$ ($|\pm 1\rangle$ and $|0'\rangle$) to a cold (hot) bath. {\bf c,} Work, in the form of stimulated emission of MW radiation, can be extracted by resonantly driving of the $|+1\rangle\leftrightarrow|0\rangle$ transition.  Note that the $|-1\rangle$ state does not contribute to the work extraction.} 
\end{centering}
\end{figure}

The specific experimental scheme we use is presented in Fig.~\ref{fig2}. Under an axial magnetic field of 0.2~T, the state $|+1\rangle$ is lowered below the state $|0\rangle$ by the Zeeman interaction, as shown in Fig.~\ref{fig2}a. It is possible to excite the  \mbox{$|+1\rangle\leftrightarrow|0\rangle$} transition with a MW field without exciting the  \mbox{$|-1\rangle\leftrightarrow|0\rangle$} transition, due to the difference in their resonance frequencies. As shown in Fig.~\ref{fig2}b, optical excitation and subsequent non-radiative decay through $|0'\rangle$ result in population inversion between \mbox{$|+1\rangle$} and \mbox{$|0\rangle$}. This enables the extraction of work, in the form of stimulated emission of MW radiation, upon resonant MW driving of the \mbox{$|+1\rangle\leftrightarrow|0\rangle$} transition (Fig.~\ref{fig2}c). Indeed, this specific system has been recently considered for the production of a chip-scale, room-temperature maser~\cite{NVMaserProposal,NV_maser}. The thermal interaction, induced by laser excitation, and the coupling to the work reservoir, resulting from resonant MW driving, can either be interlaced or both be on continuously, which correspondingly implements either a two-stroke engine or a continuous engine. 
\begin{figure}[H]
\begin{centering}
\includegraphics[width=0.65\textwidth]{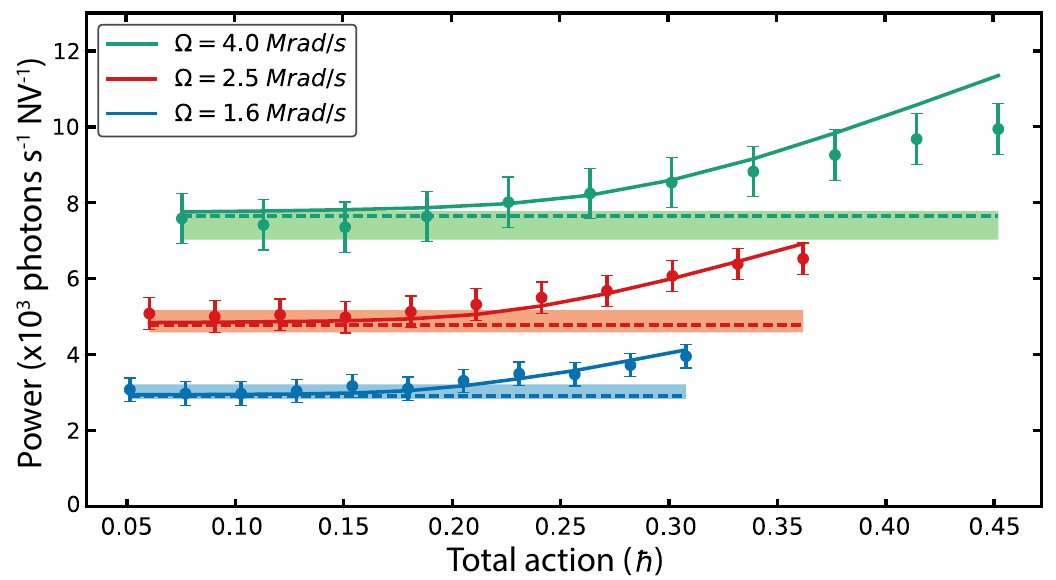}
\caption{\label{fig3} \textbf{Quantum heat machine equivalence (QHME) of two-stroke and continuous engine types.} The data points present the power output of the two-stroke engine as a function of the action per cycle (varied by changing the cycle time), for a variety of peak Rabi frequencies. The continuous output powers (together with their uncertainty) are indicated by the shaded regions. The theory predictions for the continuous (two-stroke) engine are given by the dotted (solid) line. It can be seen that, for each Rabi frequency, the output of the two-stroke engine has the continuous engine output as its zero action asymptote. }
\end{centering}
\end{figure}

It can be shown (see SI), that the average output power of the engine (per NV$^-$ centre) is proportional to the relative average change in the fluorescence intensity upon MW driving, $\langle P\rangle=\kappa(\Gamma)\hbar\omega_{10}\frac{\langle\Delta F\rangle}{\langle F_0\rangle}$, where $\langle\Delta F\rangle$ is the MW-induced change in the average fluorescence intensity, $\langle F_0\rangle$ is the average fluorescence intensity in the absence of MW driving, $\hbar\omega_{10}$ is the energy of a MW photon emitted in the transition between the two lower ground states, and $\kappa(\Gamma)$ is a rate which, for a fixed duty-cycle, depends only on the optical excitation rate, $\Gamma$. In Fig.~\ref{fig3}, the power output of the heat engine, working in two-stroke mode, is presented as a function of the engine action per cycle, varied by changing the cycle time, for a thermal bath coupling rate of $\gamma_{th}=0.41\pm0.02$~MHz and several values of the peak Rabi frequency (symbols). The duty cycle was fixed at $d=1/3$. 
In addition,  Fig.~\ref{fig3} shows the output of the continuous engine (shaded area with width signifying the measurement error. See SI), with the same mean Rabi frequencies and optical pumping rates as for the two-stroke engine. 
The convergence in performance constitute the first experimental verification of QHME. The theoretical predictions for the output power of the two-stroke (continuous) engine are presented in Fig.~\ref{fig3} by the solid (dashed) lines 
(see SI).

\begin{figure}[H]
\begin{centering}
\includegraphics[width=0.65\linewidth]{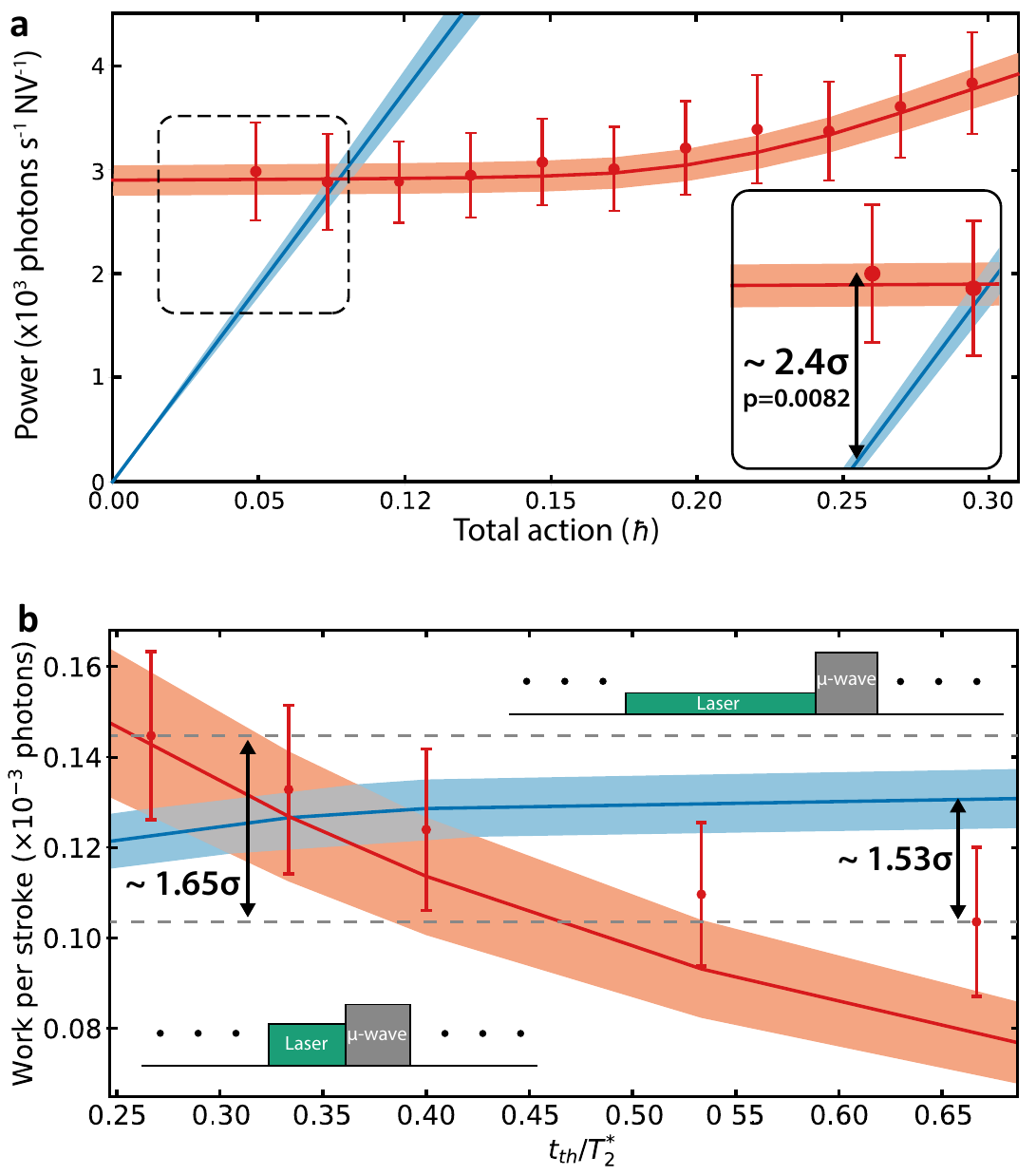}
\caption{\label{fig4} \textbf{Quantum thermal signatures.} {\bf a,} Beating the stochastic bound: The symbols show the power output of the two-stroke engine vs. the action per cycle. The solid blue line represents the stochastic bound calculated for the work strokes used in the experiment, whilst the red line is the power predicted with the theory. As indicated in the figure, for the lowest action, the measured power is four standard deviations \emph{above} the stochastic bound. {\bf b,} Work output vs. decoherence. The work per cycle of the two-stroke engine is presented vs. the thermal stroke duration (in units of the dephasing time $T_2^*=75$~ns). The dots are the measured data while the line presents the theory prediction. The work stroke length and Rabi frequency are fixed, whilst the optical excitation rate is adjusted to keep the population-transfer-related thermal action constant. The insets schematically depict cycles with a short (left) and a long (right) thermal stroke. The measured work output decreases due to the increased decoherence during the longer thermal strokes, to well below the stochastic bound.}
\end{centering}
\end{figure}

In order to measure a QTS in a quantum heat engine, we first determine an upper power bound for a fully stochastic engine, which is equivalent to the quantum engine in all respects (energy levels, coupling rates and stroke durations), bar the presence of coherence at the end of the thermal stroke. Since the power of a fully stochastic engine is constrained to lie below this bound, any measured output exceeding the bound is indicative of quantum effects.
For the case of a two-stroke engine, it can be shown that the stochastic bound is given in the small action regime by\cite{EquivPRX} $P_{stoch}  \leq \frac{1}{4}\hbar \omega_{10}d^2 \Omega^2 \tau_{cyc}$ (see SI). Fig.~\ref{fig4}a presents the measured power output of the two-stroke engine vs. the action per cycle (varied by changing the cycle time), for \mbox{$\Omega=1.6\pm 0.05$~Mrad/s}, $\gamma_{th}=0.41\pm0.02$~MHz, and $d=1/3$, along with the corresponding stochastic bound (blue line). It is clearly seen that for the smallest action applied (dashed frame, enlarged in the inset), the bound is violated by \emph{2.4} standard deviations, corresponding to a p value of 0.0082 (see SI). This constitutes a clear QTS.

We also study the work output as the coherence of the system is reduced. Fig.~\ref{fig4}b presents the work per cycle in the two-stroke engine, where the work stroke duration is fixed at 10 ns, the Rabi frequency is fixed at 1.6~MRad/s, and the thermal stroke contains a fixed population-transfer action, but is of variable duration. The total action per cycle excluding pure dephasing is fixed at 0.05$\hbar$. Thus, by changing the length of the work stroke, only the pure-dephasing related action is increased, due mostly to inhomogeneous dephasing ($T_2^*\sim75$ ns), 
enabling the examination of the dependence of the output work on the coherence of the system. The insets in Fig.~\ref{fig4}b show schemes of this cycle for short and long thermal strokes. It is clearly seen that the output work per cycle decreases as the thermal stroke duration is increased, and drops below the stochastic bound (blue line). The bound, taking into account experimental imperfections, increases slightly at long thermal strokes. The slight discrepancy between theory and measurement for long thermal strokes might be attributed either to homogeneous dephasing or to charging effects, both neglected in the present theory (see SI).
These measurements clearly demonstrate that coherence in microscopic heat engines can be performance enhancing. 

In conclusion, we have used an ensemble of NV$^-$ centres in diamond for implementing different types of quantum heat engines. We have used these to demonstrate the equivalence of the power output of two different engine types, continuous and two-stroke, for small actions. Additionally, we have shown that, for very small actions, the engines produce more power than their classical counterparts, significantly exceeding the stochastic power bound. These measurements constitute the first observation of quantum thermodynamic effects in heat machines. We hope that this work will motivate further research along at least three lines: 1) Demonstration of quantum effects in other physical realisations of heat machines such as superconducting circuits\cite{PekolaSCengine,campisi2014FT_SolidStateExp} and ion traps\cite{rossnagelIonEngExp,Sing3ionEng2017}.  2) Theoretical search of quantum thermal signatures in heat machines based on other quantum agents such as entanglement\cite{brunner14} and quantum discord. 3) Application to the design and development of novel devices such as room-temperature masers\cite{NVMaserProposal,NV_maser}. We further hope that this work will be of interest to other research areas concerned with the role of quantum coherence in the enhancement of work extraction by microscopic heat engines, such as the study of photosynthesis\cite{quant_photosynthesis} and the development of solar cells.\\

\emph{Acknowledgements:} The authors would like to thank R. Fischer and P. London for their help with sample preparation.\\

\emph{Funding:} This work was supported by the UK EPSRC (EP/J000051/1), the US AFOSR EOARD (FA8655-09-1-3020), EU IP SIQS (600645), EU COST (Action MP1209), and an EU Marie Curie Fellowship (IEF-2013-627372 to EP). PML acknowledges a European Union Horizon 2020 Research and Innovation Framework Programme Marie Curie individual fellowship, Grant Agreement No. 705278. IAW acknowledges an ERC Advanced Grant (MOQUACINO) and an EPSRC Programme Grant (EP/K034480/1).\\

\emph{Author Contributions:}  E.P. and R.U. conceived the project. J.K. performed the experiments, the data analysis, and the theoretical calculations. J.N.B., P.M.L., C.W., K.T.K., D.J.S., and E.P. contributed to various aspects of the experiment and data analysis. E.P., J.N., and I.A.W. supervised the project. J.K., R.U., and E.P. wrote the manuscript, with contributions from all authors.\\

 \emph{Competing Interests:}  The authors declare that they have no
competing financial interests.\\

 \emph{Correspondence:}  Correspondence and requests for materials
should be addressed to either E.P. (email: eilon.poem@weizmann.ac.il), J. K. (email: james.klatzow@physics.ox.ac.uk), or I.A.W. (email: ian.walmsley@physics.ox.ac.uk).
\clearpage

\renewcommand\theequation{S\arabic{equation}}
\setcounter{equation}{0}  
\renewcommand\thefigure{S\arabic{figure}}
\setcounter{figure}{0}  
\renewcommand\thetable{S\arabic{table}}
\setcounter{table}{0} 
\renewcommand\thesection{S\arabic{section}}
\setcounter{section}{0}
\renewcommand\thesubsection{S\arabic{subsection}}
\setcounter{subsection}{0}  
\renewcommand\thesubsubsection{\thesubsection.\arabic{subsubsection}}

\begin{center}
\textbf{{\large Supplementary Information for ``Experimental demonstration of quantum effects in the operation of microscopic heat engines'' }}
\end{center}

\section{Sample preparation}
A type Ib high-pressure-high-temperature 3$\times$3$\times$0.5 mm, (100) diamond slab (Element-Six) with an initial nitrogen concentration of $\sim$200~ppm was electron irradiated (10$^{18}$~cm$^{-2}$) and then annealed (950$^\circ$C, 2.5~hours), to form a dense ($\sim$10$^{18}$ cm$^{-3}$) ensemble of NV$^-$ centres. The orientations of the centres are randomly distributed between all the $\left<111\right>$ directions. However, the microwave (MW) driving is resonant only with the centres oriented parallel to the magnetic field, and thus only these centres ($\sim$25\% of all centres) produce the work.

\section{The experimental setup}
\begin{figure}[H]
	\centering
	\includegraphics[width=0.55\textwidth]{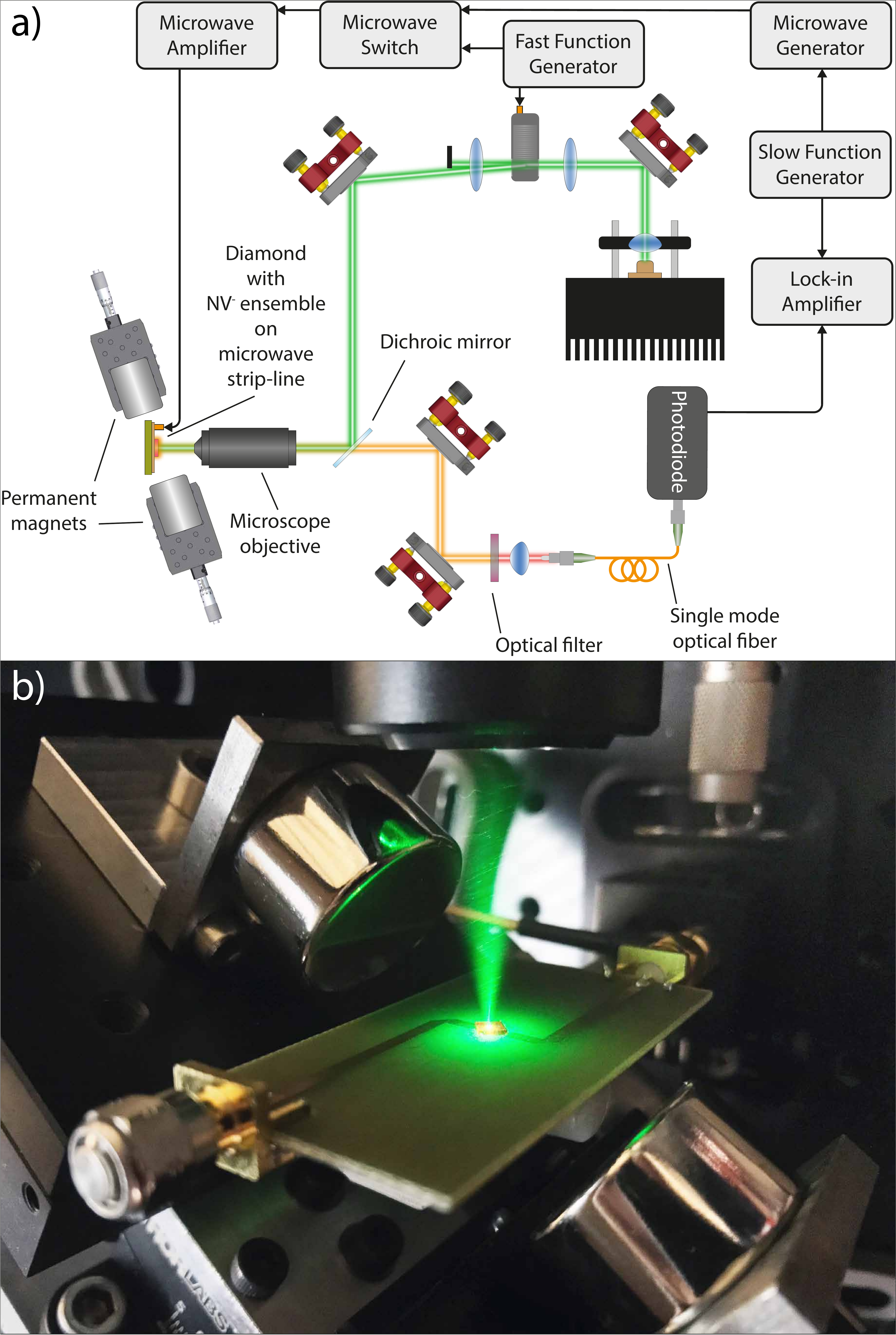}
	\caption{Experimental setup. (a) Schematic description. The fast function generator, driving the AOM and the MW switch, allows for the implementation of the two-stroke engine. Using slow amplitude modulation of the MW and lock-in detection allows for direct measurement of the net change in fluorescence induced by the operation of the engine. (b) A photograph showing the diamond sample on the MW waveguide between the two magnets. The diamond is glowing red while the green illumination is on.  }
	\label{fig2sm}
\end{figure}

The experimental system is schematically presented in Fig.~\ref{fig2sm}(a). The diamond sample was placed between two permanent magnets aligned along the $\left[111\right]$ direction. A solid-state continuous-wave laser at 532~nm was focused inside the diamond sample to a spot of (nominally) 2.2~$\mu$m in diameter by a long-working-distance objective lens (NA=0.29). An acousto-optic modulator (AOM) was used for the intensity modulation of the laser-light reaching the sample. In addition, an AC magnetic field was applied to the sample by a broad-band MW strip-line waveguide embedded in the sample holder. Just below the sample, the strip-line was narrowed down to a width of 300~$\mu$m, in order for the applied field to be both strong enough and uniform across the active volume. A photograph of the sample on the MW waveguide is presented in Fig.~\ref{fig2sm}(b). A fast MW switch (MS) between the MW generator and the MW amplifier was used for the amplitude modulation of the applied MWs.  The MS and the AOM had switching times of 1.5~ns and 12~ns, respectively, and were both simultaneously driven at $\sim$MHz rates by a fast function generator.

The fluorescence emitted from the diamond was collected by the objective lens and was imaged with a $\times$10 magnification on a single-mode optical fibre. This confocal geometry ensures the collection of light only from the central part of the laser spot, where the optical excitation rate is maximal and approximately uniform. The collected light was then detected by an avalanche photo diode operating in the linear regime. 
%

To efficiently detect the change in fluorescence due to the operation of the engine, that is, due to the MW driving, lock-in detection was used, where the output of the MW generator was amplitude-modulated by a $\sim$100 Hz square wave, much slower than the repetition rates of the engine, and a lock-in amplifier was locked to this slow modulation. 

\section{Spectroscopy and calibrations}\label{SpecCalib}
First, the distance between the magnets, as well as their mean position with respect to the sample were adjusted, such that the resonance frequency was $2\pi\cdot2600$ MHz. During this process, we observed an anti-crossing between $\left|0 \right>$ and $\left| -1 \right>$. The angle of the magnetic field was adjusted to minimise the gap at the anti-crossing point, from which we were able to infer the field angle relative to the NV axis. In our case, the angle was determined to be $0.6^{\circ}$. The transition was found to be inhomogeneously broadened with minimal full-width at half maximum of $2\pi\cdot\left(7 \pm 0.7\right)$ MHz, determined by a Gaussian fit. This implies a maximal ensemble coherence time ($1/e$) of $75 \pm 7.5$ ns. This was also confirmed by directly measuring the decoherence time using Ramsay interferometry, which yielded a similar result. Then, by measuring the change in fluorescence as a function of the MW pulse length, the Rabi-frequency was measured for different MW intensities. It was found that the ratio between the Rabi frequency and the square root of the applied MW power (after the amplifier) was $2\pi\cdot\left(244\pm 2\right)$~kHz/$\sqrt{\mbox{mW}}$. The maximum MW power available after the amplifier was 3000~mW.

\begin{figure}[H]
	\centering
	\includegraphics[width=0.6\textwidth]{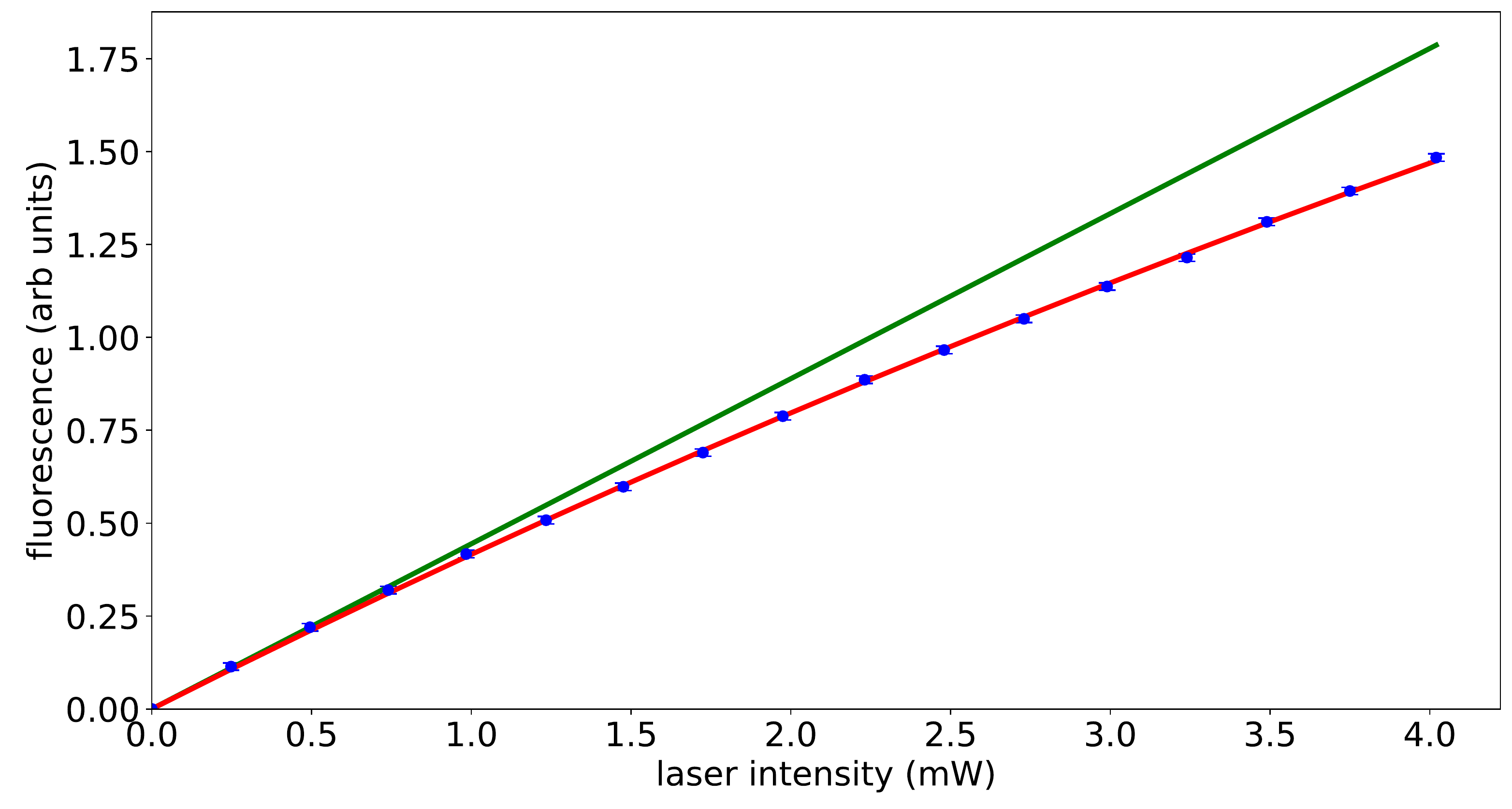}
	\caption{ Measured (symbols) and fitted (red line) fluorescence saturation curves. Using the rate equation model described in Sec.~\ref{RateEq}, the ratio between the optical excitation rate, $ \Gamma$, on the light intensity, $I_{\text{laser}}$, was determined. The green line is the tangent at zero laser intensity, showing that the saturation curve significantly deviates from linear behaviour.}
	\label{fig5}
\end{figure}

To determine the dependence of the optical excitation rate, $\Gamma$, on the laser power in our set-up, we scanned the laser power whilst measuring the total fluorescence. As can be seen in Fig.~\ref{fig5}, one obtains a curve which is linear for small laser powers, but then deviates from linear dependence. We used a rate equation model (see Sec.~\ref{RateEq}  below) 
to fit to the data using the ratio between the laser power and $\Gamma$ as a fitting parameter (together with the overall coefficient for the fluorescence). All the other parameters are known from previous, independent measurements~\cite{Tetienne_NV_mag_rates_NJP}. The obtained ratio between $\Gamma$ and the laser power measured before the objective lens was $r=436\pm25$ kHz/mW. Note that this number depends on the position in the sample of the focus of the beam, most probably due to absorption in the sample. The maximum optical power available (as measured before the objective lens) was 4.0 mW. One can compare this result with a simple calculation using the known absorption cross-section of an NV$^-$ centre at 532 nm ($\varsigma=(3.1\pm0.8)\times10^{-17}$~cm$^2$~\cite{NVcrosssection}), \mbox{$r=\varsigma/(A\varepsilon_L)$}, where $A$ is the laser spot area and $\varepsilon_L$ is the energy of one laser photon. When taking into account the measured transmission of the objective lens (81\%), the Fresnel transmission of the diamond surface (83\%), and absorption due to propagation to the middle of the sample through a dense ensemble of 10$^{18}$~cm$^{-3}$ NV$^-$ centers (45\% transmission), one finds that the focal spot diameter that yields the measured value of $r$ is $2.7\pm0.4$~$\mu$m, in good agreement with the nominal value of $2.2~\mu$m predicted for a perfect 0.29 NA objective lens. The small discrepancy might be due to deviations of the exact position of the focal spot or the exact NV$^-$ density from the estimated values used above, or due to aberrations. 

\section{Rate equation model}\label{RateEq}
The spin-state-dependent fluorescence intensity exhibited by NV$^-$ centres~\cite{NV_Review} was a part of our initial motivation for using NV$^-$ centres to implement a quantum heat engine. In this section we show how this property, quantified using a rate equation model, allows us to take all the measurements required for the experiment. 

The fluorescence spectrum for the optical excitation intensities used in this experiment contained less than 1\% neutral NV (NV$^0$) emission. Furthermore, the charging rates for these excitation intensities are expected to be on the order of $\sim1\mu$s~\cite{NV_charging_Aslam_2013}, longer than all the other characteristic times in the system (see below). We therefore neglect optical charging effects (transitions from NV$^-$ to NV$^0$ and back), and focus on the dynamics within the NV$^-$ states. 

\begin{figure}[H]
	\centering
	\includegraphics[width=0.4\textwidth]{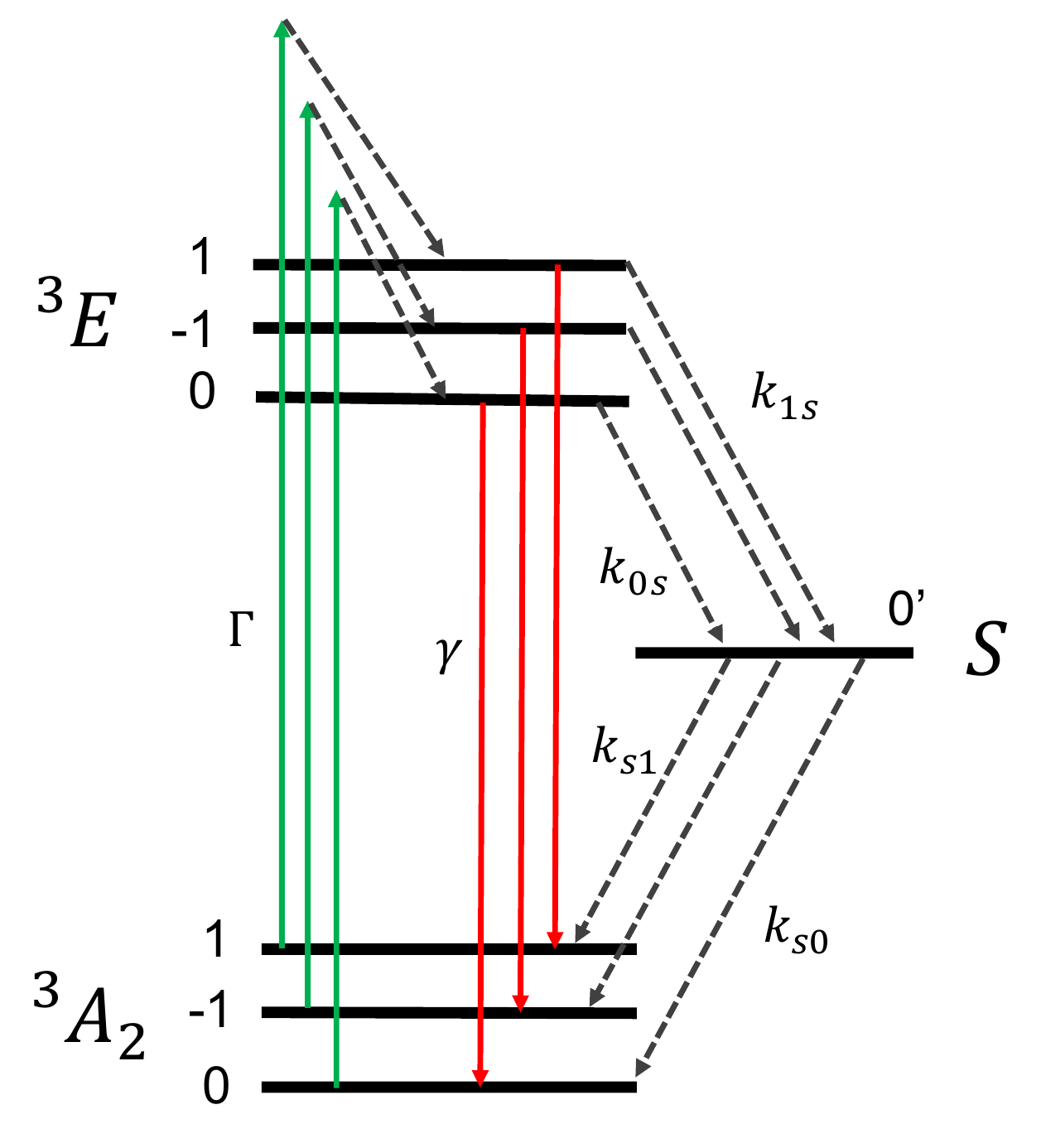}
	\caption{The simplified NV$^-$ level system used for the calculations. Horizontal lines represent levels, solid arrows represent allowed optical transitions, and dashed arrows represent non-optical transitions. G is the ground-state triplet, E represents the excited-state triplet, and S represents the three singlets (see text).}
	\label{fig8}
\end{figure}

The NV$^-$ centre consists of a ground state spin triplet, denoted by $^{3}A_2 $, two excited spin triplets, $^{3}E$, as well as three intermediate singlets, $^{1}E_{1,2}$ and $^{1}A$. Of the various interactions that lift the degeneracy of the sub-levels in the ground and excited states, orbital averaging at room temperature results in only the axial spin-spin interaction remaining \cite{time_ave_excited}; this allows us to treat the two excited state triplets as a single triplet. Additionally, the upper of the three singlets, $^{1}A_1$, decays directly into the lower pair, $^{1}E$, and has so short a lifetime ($<1$~ns) as to allow us to treat the three singlets as an effective single state . A diagram of this structure, together with the allowed transitions is shown in Fig.~\ref{fig8}.

We define the rate matrix,$\mathcal{R}$, to be that whose elements, $\mathcal{R}_{ij}$, are rates of transfer from the $i$\textsuperscript{th} to $j$\textsuperscript{th} level. Then, in the basis $\left\{ G_0 , G_{-1}, G_{1}, E_0 , E_{-1}, E_{1}, S\right\}$ (see Fig.~\ref{fig8}), $\mathcal{R}$ is given by,

$$ \mathcal{R} = \left[
\begin{array}{cccc}
	0_{3\times3}  &   \Gamma \; \mathbb{I}_{3\times3} & 0_{3 \times 1} \\
	  \gamma \; \mathbb{I}_{3\times3} & 0_{3 \times 3} &  v_{3 \times 1} \\
 	u_{1 \times 3} &  0_{1\times3} & 0_{1\times 1}\\
\end{array}
\right].$$\\

Here each of the matrix elements represents a block (with size given by the subscript),  the vectors $v$ and $u$ are given respectively by \( \left( k_{0s} \;  k_{1s} \;  k_{1s} \right)^T \) and  \( \left( k_{s0} \; \frac{1}{2} k_{s1} \; \frac{1}{2} k_{s1} \right) \), and the known~\cite{Tetienne_NV_mag_rates_NJP} rate constants are given in Table~\ref{table1}. 
 
\begin{table}[H]
\caption{\label{table1} Spontaneous decay rates between NV levels used in the analysis. Taken from Ref.~\onlinecite{Tetienne_NV_mag_rates_NJP} (averaged over all NV$^-$ orientations).}
\begin{center}
\begin{tabular}{c c}
& rate (MHz)\\ 
\hline
 $\gamma$ & $65.9 \pm 1.9$  \\  
 $k_{1s}$ & $53.3 \pm 2.5$ \\  
 $k_{0s} $ & $7.9 \pm 1.4$ \\     
 $k_{s0}$ & $0.98 \pm 0.31$ \\  
 $k_{s1}$ & $0.73 \pm 0.11$ \\
 \hline
 \end{tabular}
\end{center}
\end{table}

Under optical excitation the system will evolve according to,
\[ \partial_{t}\sigma = M\sigma, \]
where $\sigma$ represents the populations in vector form and the matrix $M$ is given in terms of the rate matrix, $\mathcal{R}$, by, 
\begin{equation} \label{rate_matrix}
M_{ij} = \mathcal{R}_{ji} - \delta_{ij} \sum_{k} \mathcal{R}_{ik}. 
\end{equation}

The rate matrix presented above is for the zero field case; when working under a magnetic field we need to determine the transformed version of  $\mathcal{R}$, and use this to calculate $M$. 
The Hamiltonians for the ground and excited state manifolds are given by,
\begin{align*}
&\mathcal{H}_{\text{gs}}  = D_{gs}S_{z}^2  + g \mu_B \vec{B} \cdot \vec{S}, \\
& \mathcal{H}_{\text{es}}  = D_{es}S_{z}^2  + g \mu_B \vec{B} \cdot \vec{S}, 
\end{align*}
where $\vec{B}$ is the magnetic field, $\vec{S}$ is the spin vector operator, $\vec{S}=\left(S_x,S_y,S_z\right)$, and all the other parameters are listed in Table~\ref{table2}.
\begin{table}[H]
\caption{\label{table2} Parameters of the NV$^-$ spin Hamiltonians. Taken from Ref.~\onlinecite{electronic_solution_NV}.}
\begin{center}
\begin{tabular}{lccc}
Parameter name & Symbol & Value & Units\\ 
\hline
 Lande g factor & $g$ & 2.00 & -  \\  
 Bohr magneton & $\mu_B$ & $2\pi\times14.0$ & GHz/T\\  
 Ground state spin-spin int.& $D_{gs}$ & $2\pi\times2.87/3$ & GHz \\     
 Excited state spin-spin int.& $D_{es}$ & $2\pi\times1.44/3$ & GHz \\ 
  \hline
 \end{tabular}
\end{center}
\end{table}

\begin{figure}[H]
	\centering
	\includegraphics[width=0.8\textwidth]{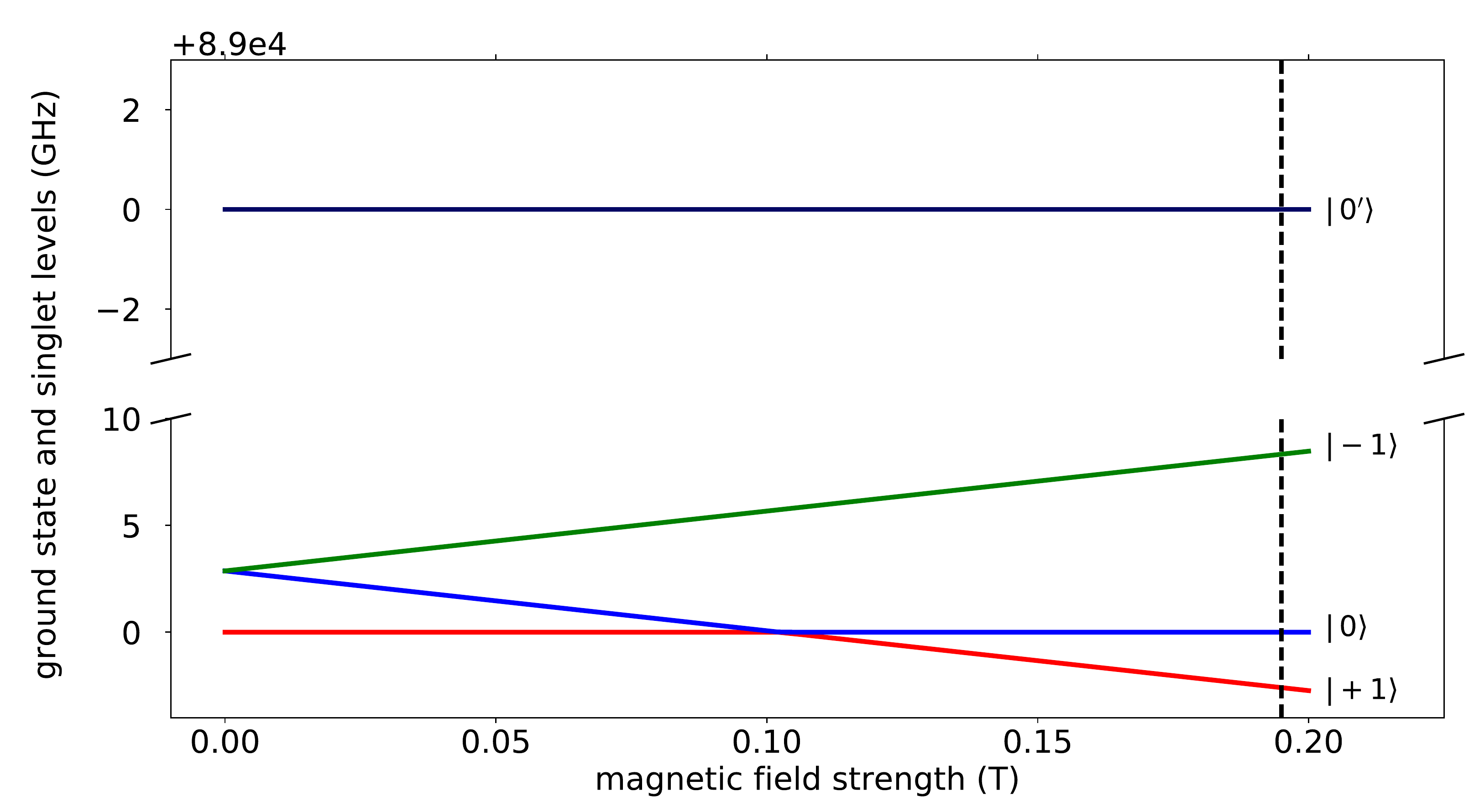}
	\caption{Singlet and ground state energy levels as a function of magnetic field strength $0.6^{\circ}$ off the NV$^-$ symmetry axis. The dashed vertical line indicates the field strength used in the experiment.}
	\label{fig14}
\end{figure}

The resulting singlet and ground-state energy levels vs. the magnetic field, for an off-axis angle of $0.6^{\circ}$, are shown in Fig.~\ref{fig14}.

\begin{figure}[H]
	\centering
	\includegraphics[width=0.75\textwidth]{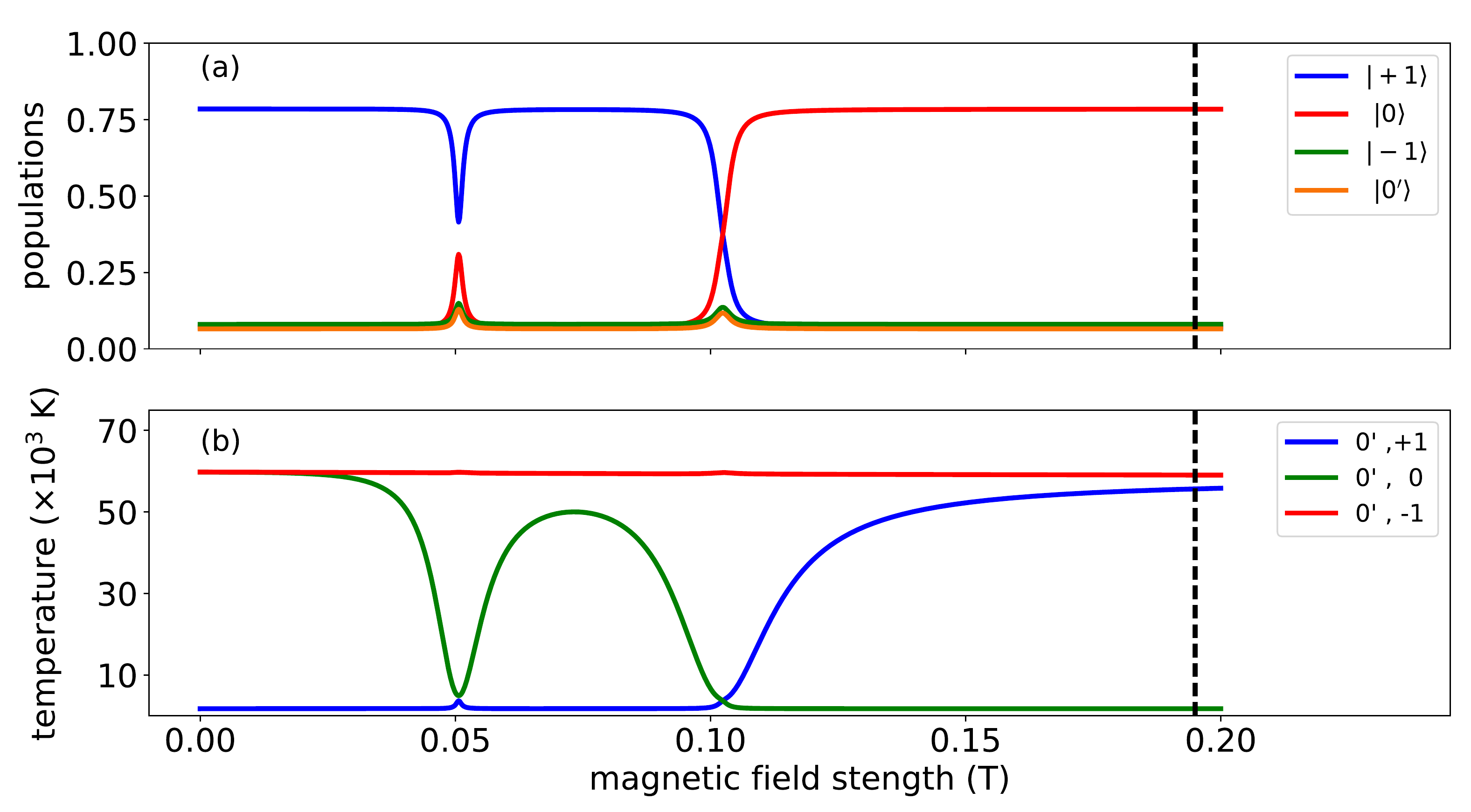}
	\caption{Steady state populations (a) and effective temperatures of the thermal reservoirs (b) as a function of the magnetic field, for the optical excitation rate used in the experiment (0.76 MHz). The legend refers to the spin state of the level at the magnetic field used in the experiment (indicated by the vertical dashed line). The ground-singlet ($^3A_2 - ^1E$) splitting used for the temperature calculation is $(89\pm 10)$~THz~\cite{Lukin_isc_dynamics_prb_15}.}
	\label{fig13}
\end{figure}

Let the unitary transformation between the zero field and final energy eigenstate be denoted by $U$. The contribution of state $\left| k\right\rangle$ to the population of the transformed state $\left| \tilde{i} \right\rangle$ is $ \left| U_{ik} \right|^2 $. Elements in the transformed rate matrix are then given by a sum over the zero-field rates, weighed by these factors:

\begin{equation*}
\mathcal{R}_{ij} \rightarrow  \sum_{k,l} \left| U_{ik} \right|^2 \left| U_{jl} \right|^2 \mathcal{R}_{kl} \implies \mathcal{R} \rightarrow \left| U \right|^2 \mathcal{R} \left| U^{T} \right|^2,
\end{equation*}

where $\left| U \right|^2$ denotes the element-wise absolute square.  In Fig.~\ref{fig13} (a) we use the matrix obtained in this manner to calculate the steady state populations of the NV centre as a function of axial magnetic field strength, for the optical excitation rate used in the experiment. The sharp changes at $\sim0.05T$ and $\sim0.1T$ correspond to anti-crossings in the excited and ground states respectively. Fig.~\ref{fig13} (b) shows the corresponding effective temperatures between the levels (as calculated using the Boltzmann factor and the known energy splittings between the ground-state triplet and the metastable singlet of $(89\pm 10)$~THz~\cite{Lukin_isc_dynamics_prb_15}). 

\section{Emulation of a thermal action}
We now show that the optical cycle in the NV$^-$ centre does indeed emulate a thermal interaction. The intuitive idea is that the optical cycle transfers population from $\left| \pm 1 \right\rangle$ to  $\left| 0 \right\rangle$ in the ground state manifold. We will formalise this and show precisely how to obtain the emulated action. The goal is to find a projection operator, $\mathcal{P}$, on the set of population vectors, together with a reduced evolution operator $L$, which satisfies the requirements for a thermal evolution operator, such that the following condition is satisfied:
\begin{equation} \label{therm_action_1}
e^{Lt} \left( \mathcal{P} \sigma \right) =  \mathcal{P}  \left( e^{Mt} \sigma \right) 
\end{equation}

This condition is a statement that the thermal operator $L$ should describe the time evolution of the reduced state, when the evolution of the whole system is generated by the optical matrix $M$. The approach we take is to treat our system as an effective 4 level system, consisting of the ground states together with the singlet, $\sigma_{red} = \left[ \; g_{0} \; , \; g_{1} \; , \; g_{-1} \; , \; s \; \right] $, which is described by the projection,
$$\mathcal{P} = \left[ \begin{array}{ccccccc} 
1 & 0 & 0 & 0 & 0 & 0 & 0 \\
0 & 1 & 0 & 0 & 0 & 0 & 0 \\
0 & 0 & 1 & 0 & 0 & 0 & 0 \\
0 & 0 & 0 & 0 & 0 & 0 & 1 \\
 \end{array}
 \right].$$

This is a justifiable choice if we can find the operator $L$ which satisfies the requirements for a thermal operator ($L_{ij} \geq 0$ for $i \neq j$, $L_{ij} \leq 0$ for $i = j$, and  $L_{ij} \leq L_{ji}$ for $i > j$ ) and Eq.~\ref{therm_action_1} to a good approximation. This will be done through an eigen-analysis of the matrix M. Using Eq.~\ref{rate_matrix}, it can be shown that $M$ is given by, 
\begin{equation*}
M = \left[ \begin{array}{ccc}
- \Gamma \; \mathbb{I}_{3 \times 3} & \gamma \; \mathbb{I}_{3 \times 3} & u \\
\Gamma \; \mathbb{I}_{3 \times 3}  & - \left( \gamma \; \mathbb{I}_{3 \times 3} + V \right) & 0_{3 \times 1}\\
0_{1 \times 3} & v & - \sum_{i=1}^3 u_i  \\
\end{array}
\right],
\end{equation*}
where $V$ is a diagonal matrix with the elements of $v$ on its diagonal.

Finding the eigenvectors of $M$ as a function of $\Gamma$ is not feasible analytically, so instead we  use perturbation theory. Ordinarily one might wish to expand about $\Gamma = 0$, however, this brings about the difficulty of dealing with the degeneracy in the ground state, which is not lifted by the first order corrections of degenerate perturbation theory. Instead we take the perturbative expansion about a value nearer the range that we use in the experiment, specifically $\Gamma = 0.5 \text{ MHz}$. The zeroth order eigenvalues are given (in units of MHz) by,
$$ \lambda^{\left(0\right)} = \left[ \; 0.0 \; , \; -0.15 \; , \; -0.22 \; , \;  -1.84 \; , \; -74.28 \; , \; -119.47 \; , \; -119.48 \;\right]$$
and have corresponding eigenvectors in the basis $\left\{ G_0, G_{-1}, G_1, E_0, E_{-1}, E_1,S\right\}$,
\begin {equation*}
\left[ \begin{array}{c} 0.99\\ 0.09\\ 0.09\\ 0.01\\ 0.00 \\ 0.00 \\ 0.06  \end{array} \right]  ,
\left[ \begin{array}{c} \; \; 0.84\\ -0.38\\ -0.38\\ \; \;0.01\\ \; \;0.00\\ \; \;0.00\\ -0.08  \end{array} \right]  ,
\left[ \begin{array}{c} \; \; 0.00 \\ \; \;0.71\\ -0.71\\ \; \;0.00\\ \; \;0.00\\ \;\; 0.00\\ \; \;0.00  \end{array} \right] ,
\left[ \begin{array}{c} -0.46 \\ -0.19\\ - 0.19\\ \;\;0.00\\ \; \;0.00\\ \; \; 0.00\\ \;\;0.85  \end{array} \right]  ,
\left[ \begin{array}{c} -0.66 \\  \; \;0.00 \\ \; \;0.00 \\ \; \; 0.74 \\ \; \;0.00 \\ \; \; 0.00\\ -0.08  \end{array} \right]  ,
\left[ \begin{array}{c}  \; \;0.00 \\  -0.10 \\  -0.43 \\ \; \;0.00 \\ \;\;0.18\\ \;\; 0.77 \\ -0.43 \end{array} \right] ,
\left[ \begin{array}{c}  \; \;0.00 \\  \; \; 0.45 \\ -0.11 \\ \; \; 0.00 \\ -0.82 \\ \; \; 0.19\\ \;\; 0.29 \end{array} \right] 
\end{equation*}

We can split these eigenvectors (and their eigenvalues) into two groups as follows: the first four eigenvectors are those which describe the changes in the 3 ground states and the singlet and for which the components corresponding to the excited states are at least an order of magnitude smaller than the other components; the other three eigenvectors have large components corresponding to the excited state triplets. This partitioning also splits the eigenvalues according to their magnitude, with the last three at least an order of magnitude greater than the first four. Note too that, if we restrict our attention to the ground state triplet together with the singlet, then the first four form a linearly independent set spanning these states. Therefore, if the system has little population in its excited states, the coefficients of the final three eigenvectors will be correspondingly small and we can write such a state in terms of the first set of eigenvectors to a good approximation. By considering the first and second order perturbative corrections, it can be shown that this is true for all $\Gamma$ used in the experiment ($\left| \Delta \Gamma \right| \leq 0.5$~MHz, where $\Delta \Gamma \equiv \Gamma - 0.5$ MHz). We now return to the problem of finding $L$. Consider Eq.~\ref{therm_action_1} -- note that this condition holds if the following holds: 
 $$\left[ L \mathcal{P} - \mathcal{P} M \right] \sigma = 0 $$

It is clearly not possible to find a $L$ which satisfies this for all possible states $\sigma$; however given that the first four eigenstates span the possible states of our system to a good approximation, we instead require that it holds on the subspace spanned by these states. We therefore require $\left[ L \mathcal{P} - \mathcal{P} M \right] \sigma_{i} = 0 $ and so $ \left[ L  -  \lambda_{i} \mathbb{I} \right] \left( \mathcal{P} \sigma_{i}\right) = 0$ for $i = 1,2,3,4$, where $\sigma_i$ ($\lambda_i$) is the $i^{th}$ eigenvector (eigenvalue) of $M$. So we see that the condition can be satisfied exactly if we require that $L$ would have eigenvectors $\left\{ \mathcal{P} \sigma_i \; |\;  i = 1,2,3,4 \right\}$ with corresponding eigenvalues $\left\{ \lambda_i \; | \; i = 1,2,3,4 \right\}$. Since there are four eigenvector/eigenvalue pairs, this completely determines $L$. Performing the calculation one finds,
 \begin{equation*}
L \left(\Gamma \right) = \; L_0 +  \Delta \Gamma L_1 + \mathcal{O}\left(\Delta \Gamma^2\right) \, ,
\end{equation*}
where,
\begin{equation*}
L_0 = \left[ \begin{array}{cccc}
-0.05 & 0 & 0 & 0.97 \\
0 & -0.22 & 0 & 0.36 \\
0 & 0 & -0.22 & 0.36 \\
0.05 & 0.22 & 0.22 & -1.71\\
\end{array}\right], \;\; \text{ and } \;\; L_1 
= \left[ \begin{array}{cccc}
-0.11 & 0 & 0 & -0.01 \\
0 & -0.45 & 0 & 0 \\
0 & 0 & -0.45 & 0 \\
0.11 & 0.45 & 0.45 & 0\\
\end{array}\right]. 
\end{equation*}

The operator $L$ generated in this way is very close to being population conserving for small $\Gamma$, but not exactly so, since some population can be transferred to the excited state manifold. Thus  we simply calculate $L$ as above, and then manually impose population conservation, modifying it by the minimal amount required to obtain the thermal evolution $L$ that we seek. 

\begin{figure}[H]
	\centering
	\includegraphics[width=0.35\textwidth]{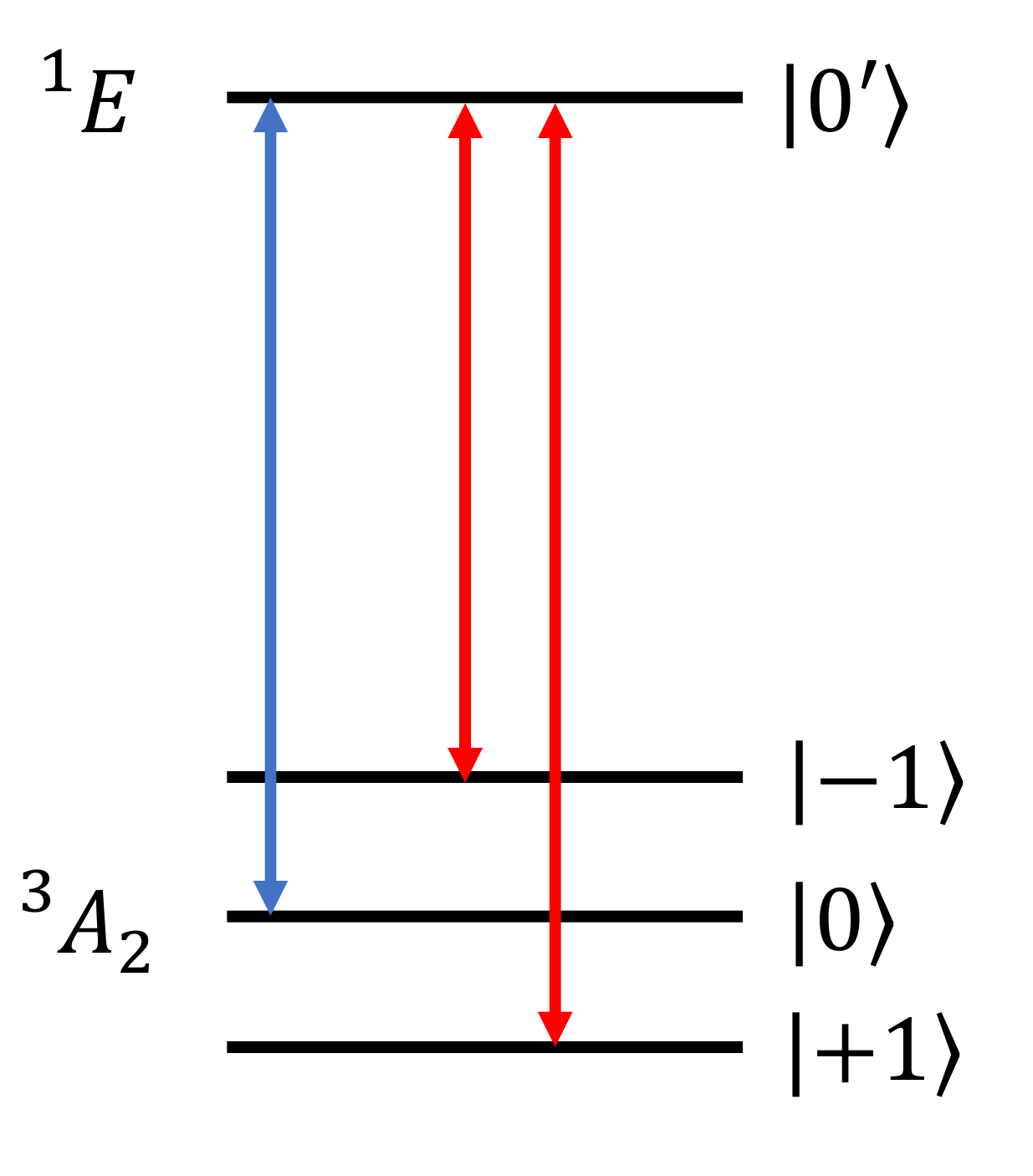}
	\caption{Schematic description of the effective thermal interaction generated by the optical excitation of the NV$^-$. Red (blue) arrows represent hot (cold) heat-bath coupling.}
	\label{fig10}
\end{figure}

We now consider what thermal reservoirs and couplings this matrix corresponds to. First notice that all the couplings are between the ground states and the singlet, with no couplings between states within the ground state triplet. Also, because the energy difference between the ground state manifold and the singlet is much greater than the differences within the ground state manifold, we can treat the effective temperature and coupling between the $\left| \pm 1 \right\rangle$ states and the singlet as being the same. Further, by examining the ratios of the rates, we see that the effective temperature between the singlet and $\left| 0 \right\rangle$ is less than that between the singlet and $\left| \pm 1 \right\rangle$. So we are left with the picture in Fig.~\ref{fig10}. 

\begin{figure}[H]
	\centering
	\includegraphics[width=0.65\textwidth]{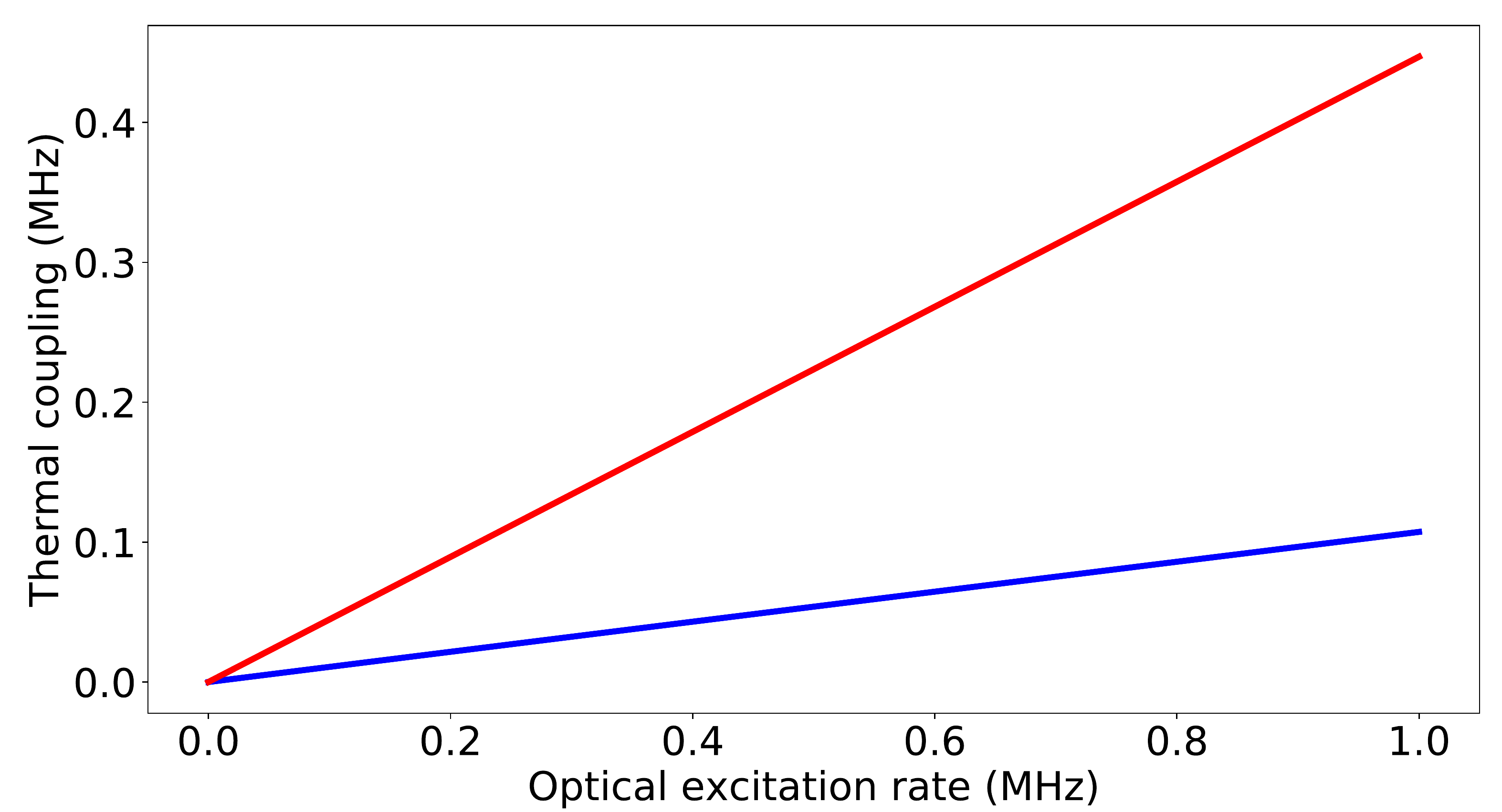}
	\caption{Effective thermal coupling rates in the emulated thermal action as a function of optical excitation rate. Red (blue) line presents the rate for the hot (cold) bath.}
	\label{fig16}
\end{figure}

Fig.~\ref{fig16} presents the effective coupling rates to the hot (red) and cold (blue) emulated heat-baths as a function of the optical excitation rate.

\begin{figure}[H]
	\centering
	\includegraphics[width=0.6\textwidth]{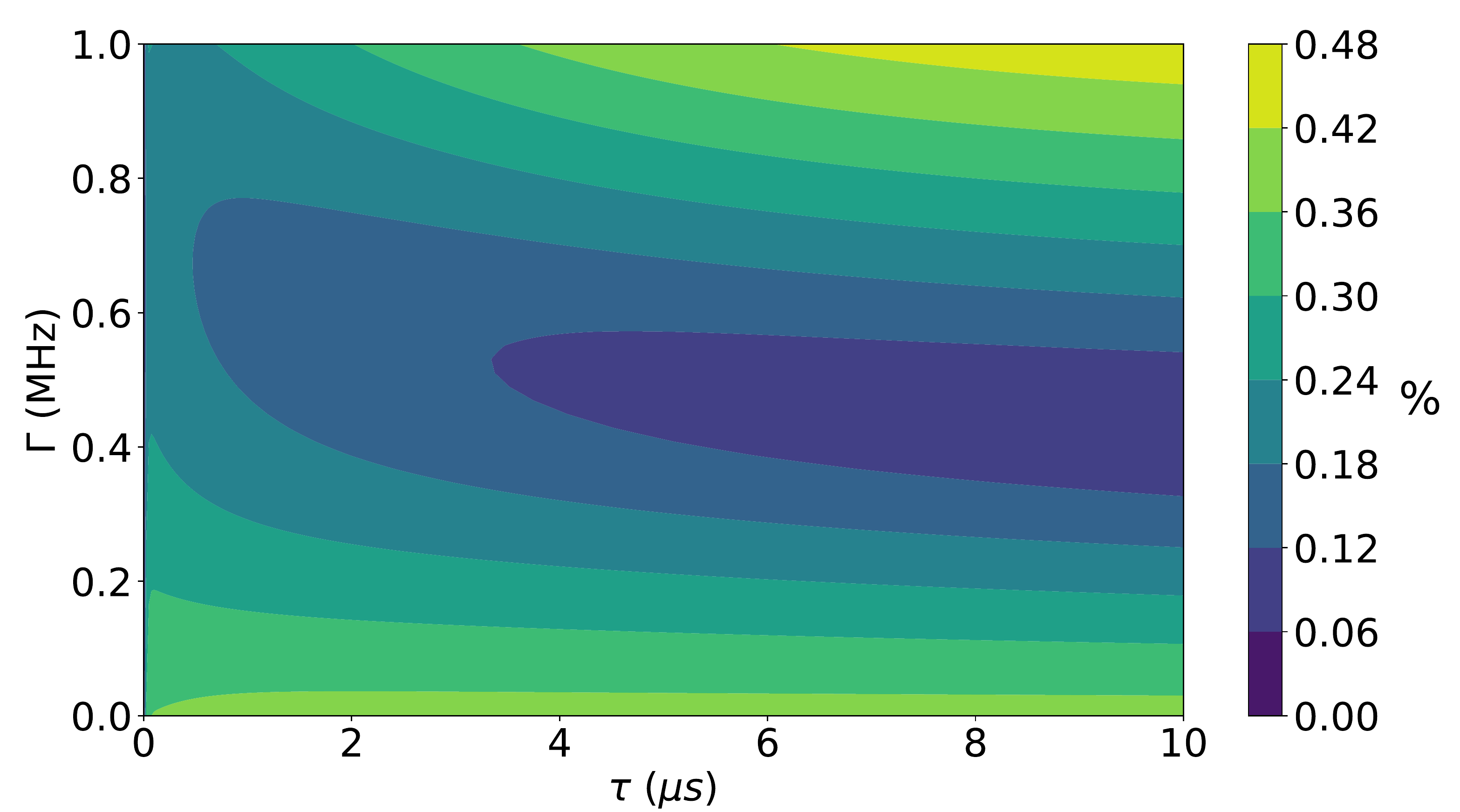}
	\caption{Difference in state evolution when using the full optical cycle as opposed to the effective thermal interaction. The horizontal (vertical) axis shows the evolution time (optical excitation rate). The scale is the percentage difference in total population.}
	\label{fig11}
\end{figure}

To demonstrate that this is a valid description, consider Fig.~\ref{fig11}, which contains a plot of $ \left\| \left[ L \mathcal{P} - \mathcal{P} M \right] \sigma_0 \right\| $ as a function of evolution time and $\Gamma$, for a starting state $\sigma_0$ with equal populations in each of the ground and singlet states and $\approx 0.5\%$ in the excited state manifold.

\begin{figure}[H]
	\centering
	\includegraphics[width=0.6\textwidth]{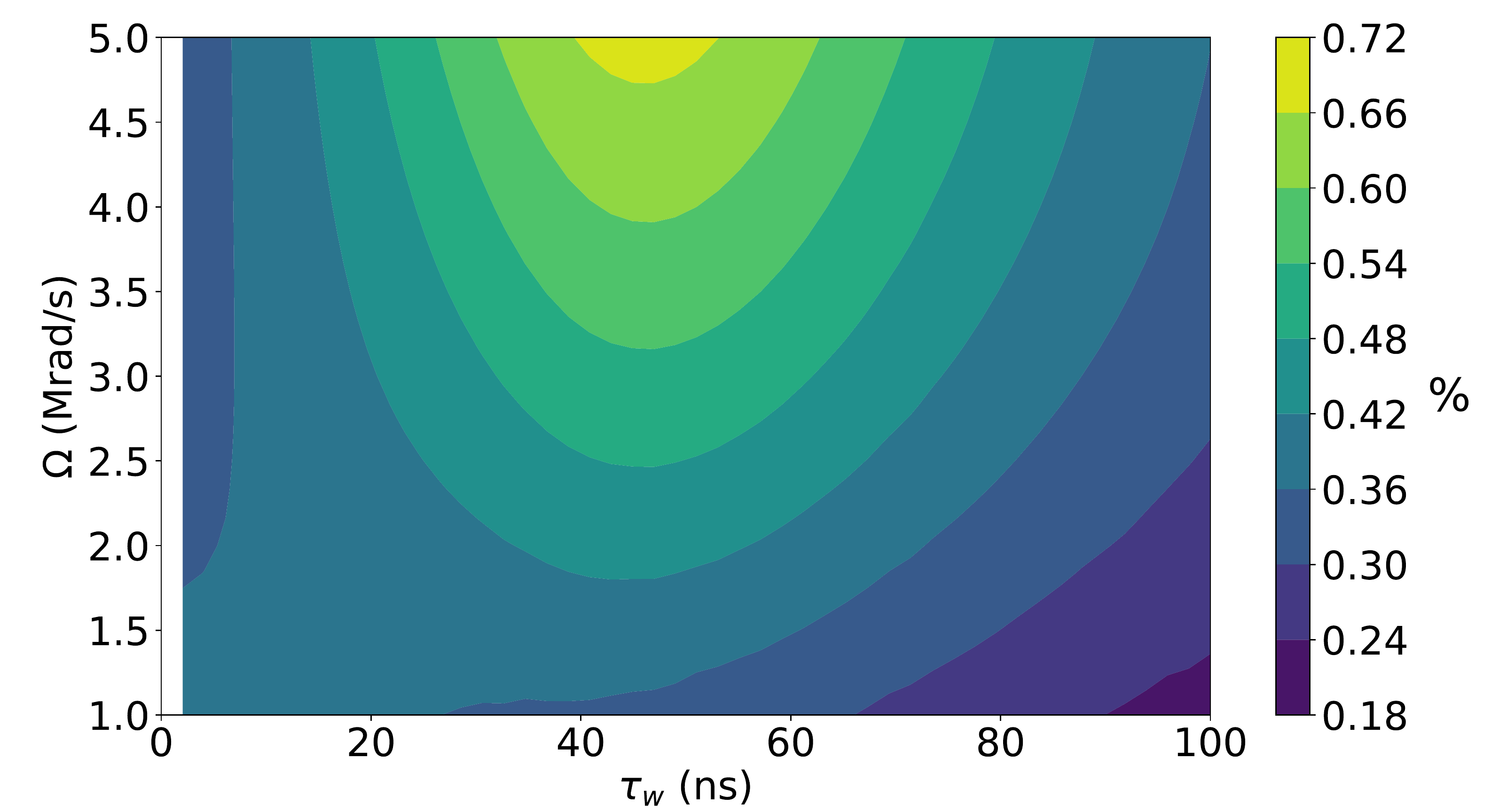}
	\caption{Difference between the two-stroke engine power output when using the full optical cycle as opposed to the effective thermal interaction. The horizontal (vertical) axis shows the work stroke duration (Rabi frequency). The parameters used are duty-cycle of d = 1/3, $\Gamma = 0.5$ MHz and an inhomogeneous broadening of $ 2\pi \times 3.0$ MHz. The scale is the percentage difference. }
	\label{fig12}
\end{figure}

We can see that the difference between the two descriptions remains below 0.5\% for the range of $\Gamma$ under consideration. We can further consider how the performance of the engine (see Sec.~\ref{engine_theory}) compares when described by the full optical matrix as opposed to the emulated thermal interaction. This is shown in Fig.~\ref{fig12}, and here too, the differences are very small within the parameter space used in the experiment. 

\section{Engine performance calculations}\label{engine_theory}
The evolution of the engine in contact with thermal reservoirs and an external field is governed by the Lindblad equation \cite{breuer,lindblad76,gorini76},
\begin{equation*}
\hbar \partial_t \rho = -i \left[ H_I , \rho \, \right] + \sum_k A_k \rho A_k^{\dagger} - \frac{1}{2} \left\{ A_k^{\dagger} A_k , \rho \, \right\},
\end{equation*}
where $H_I$ is the interaction Hamiltonian which generates unitary evolution, and $A_k$ describes the interaction with thermal baths. We can cast this into matrix operator form by vectorising the density matrix, $\rho \rightarrow \left| \rho \right\rangle$. 

$$  \hbar \partial_t \left| \rho \right\rangle = \mathcal{H} \left| \rho \right\rangle \equiv \left( -i \mathcal{H}_w + \mathcal{L} \right) \left| \rho \right\rangle, $$

$\mathcal{H}_w$ results from $H_I$ and is Hermitian, and $\mathcal{L}$ results from the remaining terms and is non-Hermitian. This formalism is known as the Liouville space (or superoperator) formalism. We can write $\mathcal{L} = \mathcal{L}_{coh} \oplus \mathcal{L}_{pop}$, where $\mathcal{L}_{pop \, (coh)}$ is the restriction of $\mathcal{L}$ to the population (coherence) subspace. In our case, we can either take $\mathcal{L}_{pop}$ to be the full rate matrix $M$, or the effective thermal operator $L$ derived in the previous section. We only have to consider the coherences $\rho_{01}$ and $\rho_{10}$ (as the MW field only couples to the $0 \leftrightarrow 1$ transition), so that $\mathcal{L}_{coh} = - \Gamma \;\mathbb{I}_{2 \times 2}$. The restriction of  $\mathcal{H}_w$ to the basis $\left\{\rho_{01}, \rho_{10}, \rho_{00}, \rho_{11}\right\}$,  is given by,

\begin{equation*}
\mathcal{H}_w \left( \Omega , \delta \right) = \frac{1}{2} \left[
\begin{array}{cccc}
-2\delta & 0 & -\Omega & \Omega \\
0 & 2\delta & \Omega  & -\Omega \\
-\Omega & \Omega & 0 &  0\\
\Omega & -\Omega & 0  & 0 \\
\end{array}
\right],
\end{equation*}
where $\Omega$ is the Rabi frequency, and $\delta$ is the detuning of the MW radiation from resonance (resulting from the inhomogeneously broadened ensemble) .

Let $\left| \, \rho_{0}(\delta)  \right\rangle$ be the steady state for a given detuning. It is given by the normalised zero eigenstate of the unitary evolution operator for a full two-stroke cycle,
$$U = e^{-i\tau_{th} \mathcal{H}_{2}} e^{-i\tau_{w} \mathcal{H}_{1}}, $$
where $\mathcal{H}_{1} = \mathcal{H}_{w}(\Omega, \delta)$, $\mathcal{H}_{2} = i\mathcal{L}(\Gamma) +  \mathcal{H}_w \left( 0 , \delta \right)$, $\tau_w$ is the work stroke duration, and $\tau_{th}$ is the thermal stroke duration. The work produced is then simply given by,
$$W \left( \delta \right) =  \left\langle H_0 \right| e^{-i\mathcal{H}_{1}t_w} - \mathbb{I}  \left| \, \rho_0 \left( \delta \right)  \right\rangle,  $$
where $\left| H_0 \right\rangle$ is the vectorised system Hamiltonian ($H_0 = \text{diag}\left[0,\omega_{10},\omega_{12}\right]$ and vectorising is achieved by reorganising the elements of $H_0$ into a column vector) and we have used the identity, $\left\langle \mathcal{O} \right\rangle = \text{Tr} \left(\mathcal{O} \rho \right) = \left\langle \mathcal{O} \right. \left|  \rho \right\rangle$. The total average output power from the ensemble is then found by integrating this over the inhomogeneous distribution, $\mathcal{N}(\delta)$, and dividing by the cycle time, $\tau_{cyc}$,
$$\langle P\rangle =\frac{1}{\tau_{cyc}} \int d \delta \; W(\delta) \mathcal{N}(\delta).$$
The detuning distribution was deduced using a Gaussian fit to the measured MW spectrum.

In the above treatment we neglected homogeneous dephasing. The main source of such a process in dense NV ensembles is spin-spin interactions between NV centres (and other paramagnetic impurities)~\cite{Lukin_magnetometer}. Its timescale is given by~\cite{Lukin_magnetometer},
$ T_2 \approx 1/\alpha n$,
where $\alpha = \mu_0g_s^2\mu_B^2/4\pi\hbar$ and $n$ is the density of NV centres. Substituting $n = 10^{18}cm^{-3}$ gives $T_2 \approx 2\ \mu$s, which is greater than the longest cycle time used in this work (180~ns) and the inhomogeneous dephasing time due to the inhomogeneous energy distribution $\mathcal{N}(\delta)$ (75~ns), allowing us to safely neglect it. 

\section{Derivation of the stochastic bound}
The existence of a stochastic bound was derived in Ref.~\onlinecite{EquivPRX}. Here we will present a brief overview of this derivation, together with a description of how the bound was calculated for our case. Such a bound must be independent of the state of a system and should also have no dependence on coherences in the system. As above we start by considering the work produced in a single cycle of the engine; however now with the coherences eliminated at the beginning and end of the cycle,
$$W= \left\langle H_0 \right| \mathcal{D} \left( e^{-i\mathcal{H}_{w}t_w} - \mathbb{I} \right) \mathcal{D} \left| \, \rho_{0} \right\rangle, $$
where $\mathcal{D}$ is a projection onto the population subspace (physically it represents a complete dephasing operation), and we include the detuning implicitly. It can be shown \cite{EquivPRX} that when one expands this expression for small actions, $s\ll\hbar$, one is left with,
$$W= \frac{\tau_w^2}{8} \left\langle H_0 \right| \mathcal{H}_w^2 \left| \, \rho_{\text{pop}} \right\rangle + \mathcal{O}\left[\left( s/\hbar \right)^4 \right],$$
where $\tau_w$ is the work stroke duration, and $\left| \, \rho_{\text{pop}} \right\rangle \equiv \mathcal{D}\left| \, \rho_{0} \right\rangle$. At this point we can simply use the known form of our Hamiltonian to show that,
\begin{align*}
W_{stoch} &= \frac{1}{4}\tau_w^2 \Omega^2 \left[ 0,\omega_{10} , \omega_{10'} \right] \left[
\begin{array}{ccc}  
\;\;1 & -1 & 0 \\
 -1 & \;\;1 & 0 \\ 
 \;\;0 & \;\;0 & 0 \\
 \end{array} \right]  \left[
\begin{array}{c}  
\rho_{11} \\ \rho_{00} \\ \rho_{0'0'}  \end{array} \right] \\
\\
&= \frac{1}{4}  \omega_{10} \tau_w^2 \Omega^2 \left( \rho_{00} -\rho_{11}\right) \leq \frac{1}{4}  \omega_{10} \tau_w^2 \Omega^2.
\end{align*}\\
Then finally the average power is given by,
$$P_{stoch}  = W_{stoch}/\tau_{cyc} \leq \frac{1}{4} \omega_{10}  d^2 \Omega^2 \tau_{cyc}, $$
where $d$ is the duty cycle.

\section{Linking the fluorescence to the engine work output}
As before, let $\sigma$ be the column vector whose elements represent the populations in the various levels and let $M$ be the optical matrix. Then given some rate, \(R(t)\), at which population is transferred from the  \(\left|0\right\rangle\) to the \(\left|1\right\rangle\) ground states, due to the microwave field, the equation describing the system is,
\begin{equation} \label{eq:1}
	\partial_{t}\sigma = M\sigma  + R(t) \nu,
\end{equation}
where  $\nu =  \left(-1\;1 \;0 \cdots 0\; 0\right)^{T}$. The rate $R$ is the quantity we wish to determine from the experimentally measured fluorescence. We proceed by considering how the inhomogeneous solution differs from the homogeneous solution (physically the difference between MW driving and no driving).

Let \( \Phi(t) \) be the fundamental matrix solution to the homogeneous version of Eq.~\ref{eq:1} (that is, a matrix which satisfies $(\partial_{t} - M)\Phi = 0$). Then it can be checked that the following is a solution to the inhomogeneous equation \ref{eq:1},
\begin{equation}\label{eq:1.5} 
	\sigma(t) = \Phi(t,0)  \sigma_{0} + \int_{0}^{t}  d\tau \; \Phi(t, \tau)  R(\tau) \nu,
\end{equation}
where $\Phi(t, \tau)  =  \Phi(t) \Phi(\tau)^{-1} $. Once past the transient period, one can take $R(t)$ to be periodic with the same period, $\tau_{cyc}$, as that of $M(t)$, as the microwaves are pulsed at the same rate as the laser pulses. This follows quite generally from Floquet theory and physical considerations. To proceed we need to find the state at the start of the cycle, \(\sigma_{0}\) , during this steady state operation. Using the periodicity of the solution we have, 
$$ \sigma_{0} = \sigma(\tau_{cyc}) =  \Phi(\tau_{cyc},0)  \sigma_{0} + \int_{0}^{\tau_{cyc}}  d\tau \; \Phi(\tau_{cyc}, \tau)  R(\tau) \nu, $$
and therefore,
\begin{equation} \label{eq:2}
\left[ \mathbb{I} - \Phi(\tau_{cyc},0) \right] \sigma_{0} = \int_{0}^{\tau_{cyc}}  d\tau  \Phi(\tau_{cyc} , \tau)  R(\tau) \nu. 
\end{equation}

We would like to solve for \(\sigma_0\), however the matrix on the left hand side is singular. Note that $\Phi(\tau_{cyc})$ is diagonalisable [it has an inverse: $\Phi(-\tau_{cyc})$], which allows us to write $ \left( \mathbb{I} - \Phi(\tau_{cyc}) \right) = U D U^{-1}$, where $D$ is diagonal. We define the pseudo-inverse,  $\left( \mathbb{I} - \Phi(\tau_{cyc}) \right)^{-} = U D^{-} U^{\dagger}$ ,  where the diagonal matrix $D^{-} $ is given by: 
\[ D^{-}_{ii} =
\begin{cases}
    1/D_{ii} \hspace{5pt}&\text{if } D_{ii} \neq 0 \\
    0 \hspace{5pt} &\text{otherwise}
\end{cases}
\]
 
Let $\mathcal{A} = \left( \mathbb{I} - \Phi(\tau_{cyc}) \right)^{-}$. A solution to Eq.~\ref{eq:2} is then given by:
$$ \tilde{\sigma}_0 =  \mathcal{A}  \int_{0}^{\tau_{cyc}}  d\tau \; \Phi(\tau_{cyc}, \tau)  R(\tau) \nu. $$

The full space of solutions to Eq.~\ref{eq:1} is then generated by $\alpha \rho_0 + \tilde{\sigma}_0$ for $\alpha \in\mathbb{R} $, where $\rho_0$ is the (unique) eigenvector of $\Phi(\tau_{cyc})$ with eigenvalue 1. We now show that $\sum_j\left[\tilde{\sigma}_0\right]_j=0$ : Suppose $\left\{ \rho_i \right\}$ are the eigenvectors of $\Phi(\tau_{cyc})$ with corresponding eigenvalues $\left\{ \lambda_i \right\}$. Then by virtue of $\Phi(\tau_{cyc}) $ being population preserving we have,

$$\sum_j \left[\Phi(\tau_{cyc})\rho_i \right]_j =\lambda_i\sum_j \left[\rho_i\right]_j=\sum_j \left[\rho_i\right]_j,  $$ 

which implies $\sum \left[\rho_i\right]_j=0 $ for $i \geq 1$. Then given a vector $\omega$, $\sum_j\omega_j =0 \implies \omega \in \text{span} \left( \rho_1, \dots , \rho_n\right) $. It is also clear that $\text{Im}\left(\mathcal{A}\right) \subseteq \text{span}\left( \rho_1, \dots , \rho_n\right)$. The unique normalised solution, $\sigma_0$, to Eq.~\ref{eq:2} is therefore simply obtained by setting $\alpha =1 $ (assuming we have normalised  $\rho_0$). Substituting this back into Eq.~\ref{eq:1.5}, we have,
$$ \sigma(t) = \Phi(t,0)  \rho_{0} +  \Phi(t,0) \hspace{5pt}   \mathcal{A}  \int_{0}^{\tau_{cyc}}  d\tau \; R(\tau) \Phi(\tau_{cyc}, \tau) \vec{\nu}   + \int_{0}^{t}  d\tau \;  R(\tau)  \,\Phi(t, \tau) \nu $$
and therefore,
\begin{equation*}
\sigma(t) - \rho(t) =   \Phi(t,0) \hspace{5pt}  \mathcal{A}   \int_{0}^{\tau_{cyc}}  d\tau \; R(\tau) \Phi(\tau_{cyc}, \tau) \vec{\nu}   + \int_{0}^{t}  d\tau \; \Phi(t, \tau)  R(\tau) \nu,
\end{equation*}
where $\rho(t)=\Phi(t,0)\rho_0$ is the state evolution when $R =0$. We define the excited state population projector to be the vector $\Omega_{E} = (0,0,0,1,1 ,1,0)$ , so that the fluorescence rate, $F(t)$, from a state \(\sigma(t)\) is proportional to $\Omega_E \cdot \sigma(t)$, where $\zeta$ is the radiative decay rate. Denote the time dependent fluorescence in the presence of MW driving by $F(t)$ and denote the fluorescence in the absence of MW driving by  $F_0(t)$. Further define,
$$ g(t,\tau) = \Omega_{E} \cdot   \Phi(t,0) \hspace{5pt}   \mathcal{A}   \; \Phi(\tau_{cyc}, \tau) \cdot \nu \hspace{10pt} \text{and} \hspace{10pt} f(t,\tau) =   \Omega_{E}\cdot  \Phi(t, \tau) \cdot \nu. $$

We can then write,
\begin{equation*}
\left[ \frac{1}{\left\langle F_0 \right\rangle \tau_{cyc}} \int_{0}^{\tau_{cyc}}  d\tau \; \Omega_{E} \cdot \rho(\tau)  \right] \times \left(F(t)  - F_0(t)\right)=   \int_{0}^{\tau_{cyc}}  d\tau \; g(t,\tau)R(\tau) + \int_{0}^{t}  d\tau \; f(t,\tau) R(\tau),  \end{equation*}
where the term in square parenthesis is the proportionality constant linking the fluorescence to the excited state population. This can be written more succinctly  as,
\begin{equation*}
\left[  \frac{1}{\tau_{cyc}} \int_{0}^{\tau_{cyc}} d\tau \; \Omega_{E} \cdot \rho(\tau)  \right] \times \frac{\Delta F(t)}{\left<F_0\right>} =   \int_{0}^{\tau_{cyc}}  d\tau \; h(t,\tau)R(\tau), 
\end{equation*}
where we have defined $\Delta F(t)=F(t)-F_0(t)$,  and 
\begin{equation*}
h(t,\tau) =
\begin{cases}
    g(t,\tau)+f(t,\tau) \hspace{5pt}&\text{if } t>\tau \\
    g(t,\tau) \hspace{5pt} &\text{otherwise.}
\end{cases}
\end{equation*}

Experimentally, we only measure  the change in the \emph{average} fluorescence,  \( \langle \Delta F \rangle \), so we must integrate \(t\) over the period to obtain,
\begin{equation*}
\left[  \frac{1}{\tau_{cyc}} \int_{0}^{\tau_{cyc}} d\tau \; \Omega_{E} \cdot \rho(\tau)  \right] \times \frac{\left<\Delta F\right>}{\left<F_0\right>} =   \dfrac{1}{\tau_{cyc}} \int_{0}^{\tau_{cyc}}  d\tau \; H(\tau)R(\tau),  
\end{equation*}
where $H(\tau) = \int_{0}^{\tau_{cyc}} dt \, h(t,\tau)$. It can be shown numerically that, for sufficiently small cycle duration and optical pumping rate, $H$ is constant to a good approximation (for the maximum stroke duration and optical intensity used in the experiment, the relative change of $H$ over the interval is $\sim 4 \times 10^{-4}$). This allows us to write,
\begin{equation}
\left< R\right> = \kappa(\Gamma ) \frac{\left<\Delta F\right>}{\left<F_0\right>}, 
\end{equation}
where,
\[\kappa(\Gamma) = \frac{1}{H\tau_{cyc}} \int_{0}^{\tau_{cyc}} d\tau \; \Omega_{E} \cdot \rho(\tau).\]

\begin{figure}[H]
	\centering
	\includegraphics[width=0.8\textwidth]{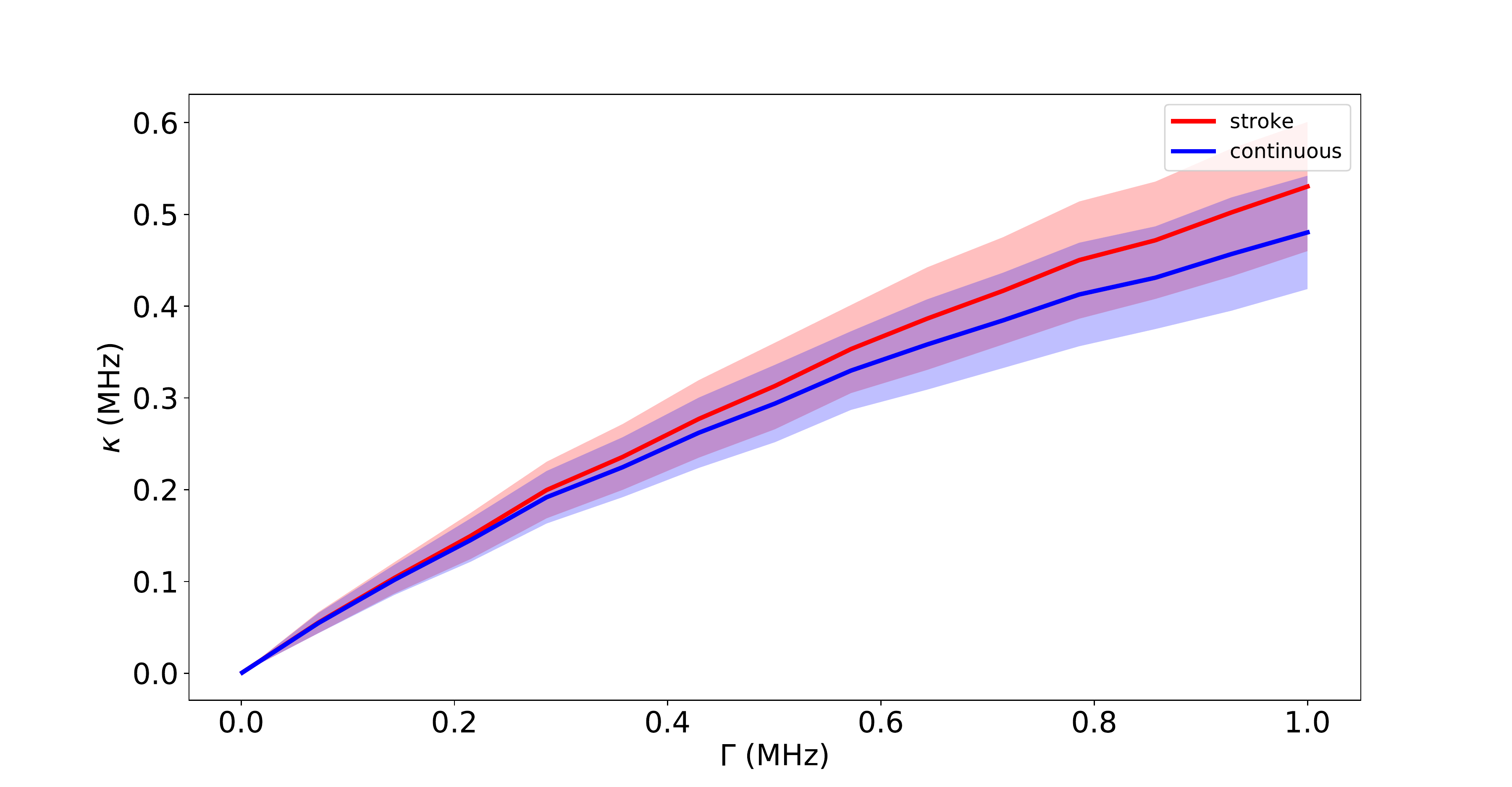}
	\caption{$\kappa(\Gamma)$ for a continuous engine (red), and two-stroke engine with a duty cycle of $d=1/3$ (blue). The shaded areas represent the error in $\kappa$ (one standard deviation).}
	\label{fig15}
\end{figure}

Note that $\kappa(\Gamma)$ does not depend on the MW driving, but only on the optical pumping rate, $\Gamma$, and the known decay rates of the system. Fig.~\ref{fig15} presents $\kappa$ vs. the optical excitation rate $\Gamma$ for both continuous and two-stroke engines, for the range of optical excitation rates used in the experiment. The error (one standard deviation) in the value of $\kappa$, estimated using the Monte-Carlo method, is presented by the shaded areas.\\

The average power is then simply given by,
\begin{equation} \label{power_output}
\langle P\rangle = \hbar \omega_{10}\left< R\right>=\hbar \omega_{10} \kappa \left( \Gamma \right) \frac{\left<\Delta F\right>}{\left<F_0\right>}, 
\end{equation}

where \( \hbar\omega_{10}\) is the energy gap between the relevant levels.

\section{Uncertainty Analysis}
In examining expression \ref{power_output} it is apparent that there are a number of contributions to the uncertainty in the final value of the power, $\langle P\rangle$. There is measurement variation in the value $\langle \Delta F \rangle$ from one measurement to the next, due to detector shot noise. The mean and standard error were simply calculated from a sample of individual measurements. The value for $\kappa$ depends on the parameters listed in table \ref{table1}, together with the value calculated for $\Gamma$, which itself depends on the former parameters. We use a Monte-Carlo simulation to propagate the uncertainties in these parameters, and the calculated uncertainty for $\Gamma$, to a final uncertainty in $\kappa$. The error in the value for $\langle F_0\rangle$ was determined to be negligible (relative error $ < 1\%$) relative to the other errors and therefore disregarded. Finally we also need to account for uncertainty in the value for the bound. This stems principally from the uncertainty in the value for the Rabi frequency. 

The quantity of interest is the certainty with which we can demonstrate $P_{measured} - P_{bound} > 0$. This requires that we use a one sided normal distribution. The null hypothesis is $P_{measured} - P_{bound} \leq 0$, whilst the alternative hypothesis is that the measured power breaks the stochastic bound. In our case the test statistics for the null outcome is $t = 2.4$, which corresponds to a p-value of 0.0082. Thus we can discard our null hypothesis and adopt the alternative hypothesis, to a significance of $<1\%$. 

\bibliographystyle{naturemag}
\bibliography{nv_heat_engine.bib}

\end{document}